\documentclass[
 amsmath,amssymb,
 aps, nofootinbib
]{revtex4-2}
 \pdfoutput=1
\usepackage{ulem}
\usepackage{graphicx}
\usepackage{amsmath}
\usepackage{amsfonts}
\usepackage{amssymb}
\usepackage{subfigure}
\usepackage{amsthm}
\usepackage{amssymb}
\usepackage{amsxtra}     
\usepackage{epsfig}
\usepackage{verbatim}
\usepackage{graphics}
\usepackage{wrapfig}
\usepackage{subfigure}
\usepackage{float}
\usepackage{morefloats}
\usepackage{xcolor}

\begin{document}
\title{Functional, Durable, and Scalable Origami-Inspired Springs}
\author{Shadi Khazaaleh\footnote{PhD Student, Engineering Division, New York University (NYU), Abu Dhabi, e-mail: smk24@nyu.edu}  \qquad Ravindra Masana\footnote{Research Associate at the Engineering Division, New York University (NYU), Abu Dhabi, e-mail: rm4829@nyu.edu} \qquad Mohammed F. Daqaq\footnote{Global Network Professor of Mechanical Engineering at NYU and NYU-Abu Dhabi  e-mail: mfd6@nyu.edu}}
\affiliation{Laboratory of Applied Nonlinear Dynamics (LAND)\\Engineering Division \\ New York University, Abu Dhabi, UAE.}

\date{\today}

\begin{abstract}
Origami has recently emerged as a platform for building functional engineering systems with versatile characteristics that targeted niche applications. One widely utilized origami-based structure is known as the Kresling origami spring (KOS), which inspired, among many other things, the design of vibration isolators, fluidic muscles, and mechanical bit memory switches. Previous demonstrations of such concepts were carried out using paper-based KOSs, which are suitable for conceptual illustrations but not for implementation in a real environment. In addition to the very low durability resulting from the high plastic deformations at the paper folds; lack of repeatability, and high variation of performance among similar samples are also inevitable. To circumvent these issues, this paper presents a novel approach for the design of a non-paper based KOS, which mimics the qualitative behavior of the paper-based KOS without compromising on durability, repeatability, and functionality. In the new design, each fundamental triangle in the paper-based KOS is replaced by an inner central rigid core and an outer flexible rubber-like frame, which are fabricated out of different visco-elastic materials using advanced 3-D printing technologies. The quasi-static behavior of the fabricated springs was assessed under both compressive and tensile loads. It is shown that KOSs with linear, softening, hardening, mono- and bi-stable restoring force behavior can be fabricated using the proposed design by simple changes to the geometric design parameters, further pointing to the vast range of potential applications of such springs.
\end{abstract}

\maketitle


\section{Introduction}
The word \textit{origami} is typically associated with aesthetically appealing structures folded out of colorful paper in the form of birds, flowers, and butterflies \cite{MahadevanOrigami2005,Faber1386}. Many fail to realize that, even before becoming a form of art, astounding and intricate forms of origami were adopted in nature by plants and insects over millions of years of evolution \cite{ArthurOrigamiReview2014,SchenkReview2014,CAVALLO2015538}. Today, origami has also emerged as a platform for building functional engineering systems with versatile characteristics that targeted niche applications spanning different technological fields \cite{SchenkReview2014,Sli2019_ArchitectedOrigami,Jiang2014OrigamiMetamaterials}. This includes building structures with auxeticity \cite{schenk2013geometry}, multistability \cite{waitukaitis2015origami}, and programmable stiffness \cite{Silverberg2014,fang2018programmable}. 

Origami-inspired structures are typically made by folding paper or flat sheets of foldable materials following certain patterns that result in different shapes and  functionalities. For example, the \textit{Miura-Ori} pattern has been used to construct three-dimensional deployable structures that have been extensively studied and utilized in applications ranging from space exploration~\cite{MiuraOrigami1985,nishiyama2012miura,miura2009science,miura19852}, deformable electronics \cite{SongOrigamiBatteries2014,Rui2014}, artificial muscles \cite{OrigamiMuscles2017}, and reprogrammable mechanical metamaterials \cite{Silverberg2014,Sengupta2018,Sli2019_ArchitectedOrigami}.  The \textit{Kresling} pattern has also garnered significant attention within the origami-inspired engineering community \cite{Butler2016,Kidambi2020,kaufmann2020harnessing,Kidambi2020Kresling}. It has inspired the design of flexible tunable antennas \cite{Yao_OrigamiAntenna2014_2}, robot manipulators \cite{Twistedtowerrobot}, wave guides \cite{yasuda2019origami}, selectively-collapsible structures \cite{ZhaiKresling2018}, high flexural strength soda cans \cite{HanOrigamiCan2004}, and crawling and peristaltic robots \cite{BHOVAD2019100552,TawfickKreslingRobot2017}.

\begin{figure*}[htb!]
		\centering 
		\includegraphics[width=12.0cm]{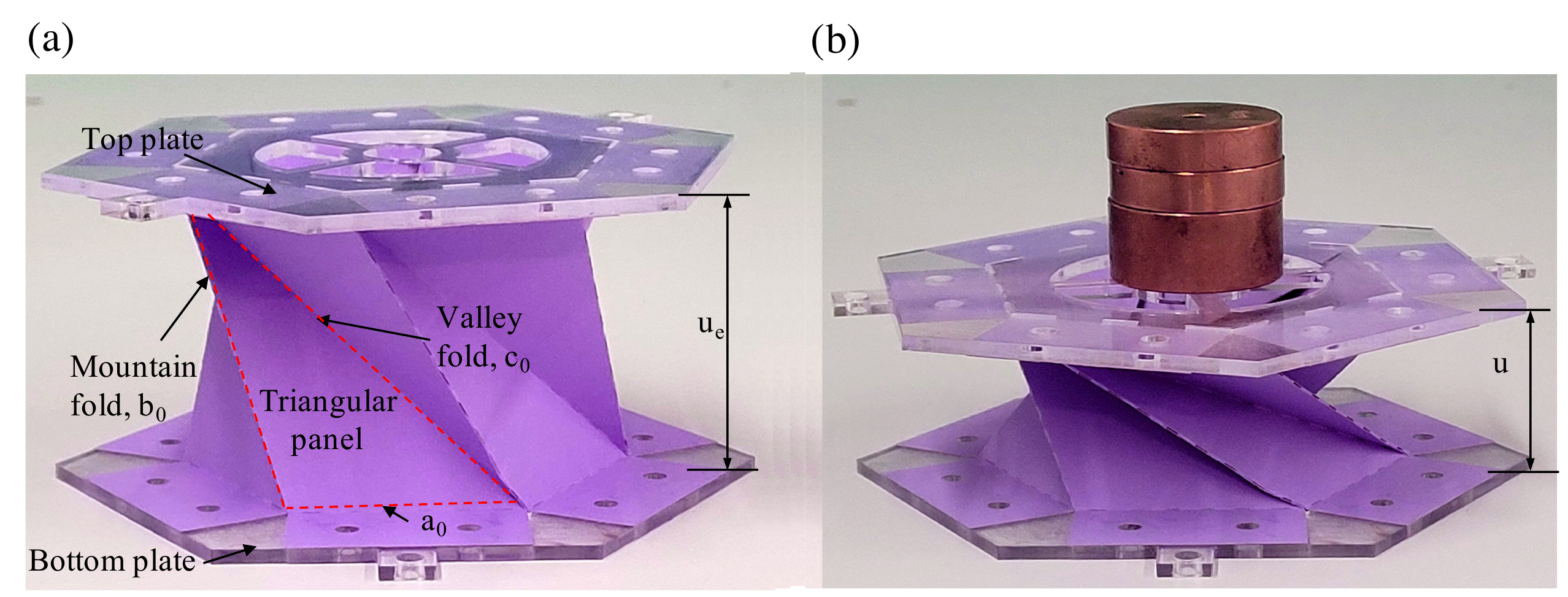}
		\caption{Paper-based KOS. (a) Unloaded, and (b) under a compressive axial load. } \label{Scheme}
	\end{figure*}

One widely utilized structure based on the Kresling pattern is known as the Kresling origami spring (KOS), which inspired the design of vibration isolators \cite{ishida2017design},  fluidic muscles \cite{OrigamiMuscles2017}, and mechanical bit memory switches \cite{YasudaKresling2017,MasanaKIMS}. A KOS is typically constructed by folding paper following a particular procedure that involves segmenting a flat sheet of paper into triangles and folding it along the edges to form valley and mountain folds (see Appendix A for details). The result is a cylindrical bellow-type structure consisting of similar triangles arranged in cyclic symmetry and connected together as shown in Fig. \ref{Scheme} (a). The circular arrangement of the triangles connects each triangle in the KOS to two other triangles along two of its edges, with one forming a valley fold and another a mountain fold. Together, the third edges of the connected triangles form two parallel polygonal end planes.

When an external axial load or a torque is applied to the KOS, it stretches or compresses depending on the direction of the applied load. In the process, the two parallel polygon planes, while remaining rigid as shown in Fig. \ref{Scheme} (b), move and rotate relative to each other along and about a common centroidal axis similar to the motion of a threaded screw into a nut (watch video SI-1 in the supplementary files for the motion of the KOS under the application of a compressive load). This causes the triangular panels to deform and store the applied work in the form of strain energy. The stored energy is released upon the removal of the external loads which forces the KOS to spring back to its initial configuration, therewith providing a restoring element that forms the basis for the design of many interesting engineering structures.
 
Kresling origami springs are traditionally built out of foldable materials, mainly paper, using conventional fabrication processes which include manual folding and creasing \cite{Butler2016,yasuda2019origami,BHOVAD2019100552,TawfickKreslingRobot2017,RmasanaKIOS19}. Such materials and fabrication methods are suitable for conceptual illustrations and laboratory testing, but are nowhere suitable for implementation in a real environment. In addition to the very low durability resulting from the high plastic deformations at the folds; lack of repeatability, and high variation of performance among samples of same design are also inevitable due to human errors and variations between material samples \cite{RmasanaKIOS19}. One clever approach to increase durability and functionality was proposed by Ishida et al.~\cite{ishida2017design} and Yasuda et al. \cite{YasudaKresling2017}, who proposed removing the panels and using flexible linkages along the edges of the panels to mimic the behavior of the folds. They used this approach to develop KOSs that can act as vibration absorbers and mechanical memory bits. However, the complexity of the truss design remain one of the fundamental issues that need to be resolved before low-cost mass manufacturing of such springs becomes realizable.

In this paper, we utilize our understanding of the quasi-static behavior of paper-folded KOSs to propose a novel design which involves replacing each fundamental triangle in the paper based-KOS by an inner central rigid core and an outer flexible rubber-like frame. We use advanced 3-D printing technologies to fabricate them, and study their quasi-static behavior for different design parameters. We show that the fabricated springs mimic the qualitative behavior of the paper-based springs without compromising on durability, repeatability, and functionality.  We also show that their behavior can be qualitatively changed by simple adjustments to the geometric design parameters. We investigate the quasi-static behavior of stacked KOSs, and based on the resulting quasi-static behavior, we propose possible avenues for their application. We believe that the proposed design will open new avenues for the mass manufacturing of KOSs, which will facilitate their implementation in real engineering applications.

The rest of the paper is organized as follows: Section \ref{designfundamentals} describes the fundamental geometric design aspects of paper-based KOSs, and introduces a simplified truss model, which can be used to study their qualitative quasi-static behavior. Section \ref{maps} uses the truss model to generate a design map, which can be used to identify the set of geometric parameters that lead to mono- versus bi-stable restoring force behavior of the KOS. Section \ref{fabrication} outlines the design of the new non-paper based KOSs, and details the fabrication process. Section \ref{quasi} investigates experimentally the quasi-static behavior of the fabricated KOSs for different geometric design parameters that lead to distinctive qualitative behaviors.  Section \ref{new} studies the effect of the key geometric design parameters associated with the new design on the quasi-static behavior of the KOS. Section \ref{scaling} investigates the effect of scaling on the restoring force behavior of the KOS. Section \ref{stacking} studies the quasi-static behavior of a KOS stack consisting of either two mono- or bi-stable KOSs. Section \ref{Durability} investigates the durability of the KOS by testing its behavior after a large number of loading cycles. Finally, section \ref{conclusion} presents the key conclusions. 

\section{Design Fundamentals}
\label{designfundamentals}
\begin{figure*}[htb!]
		\centering 
		\includegraphics[width=8cm]{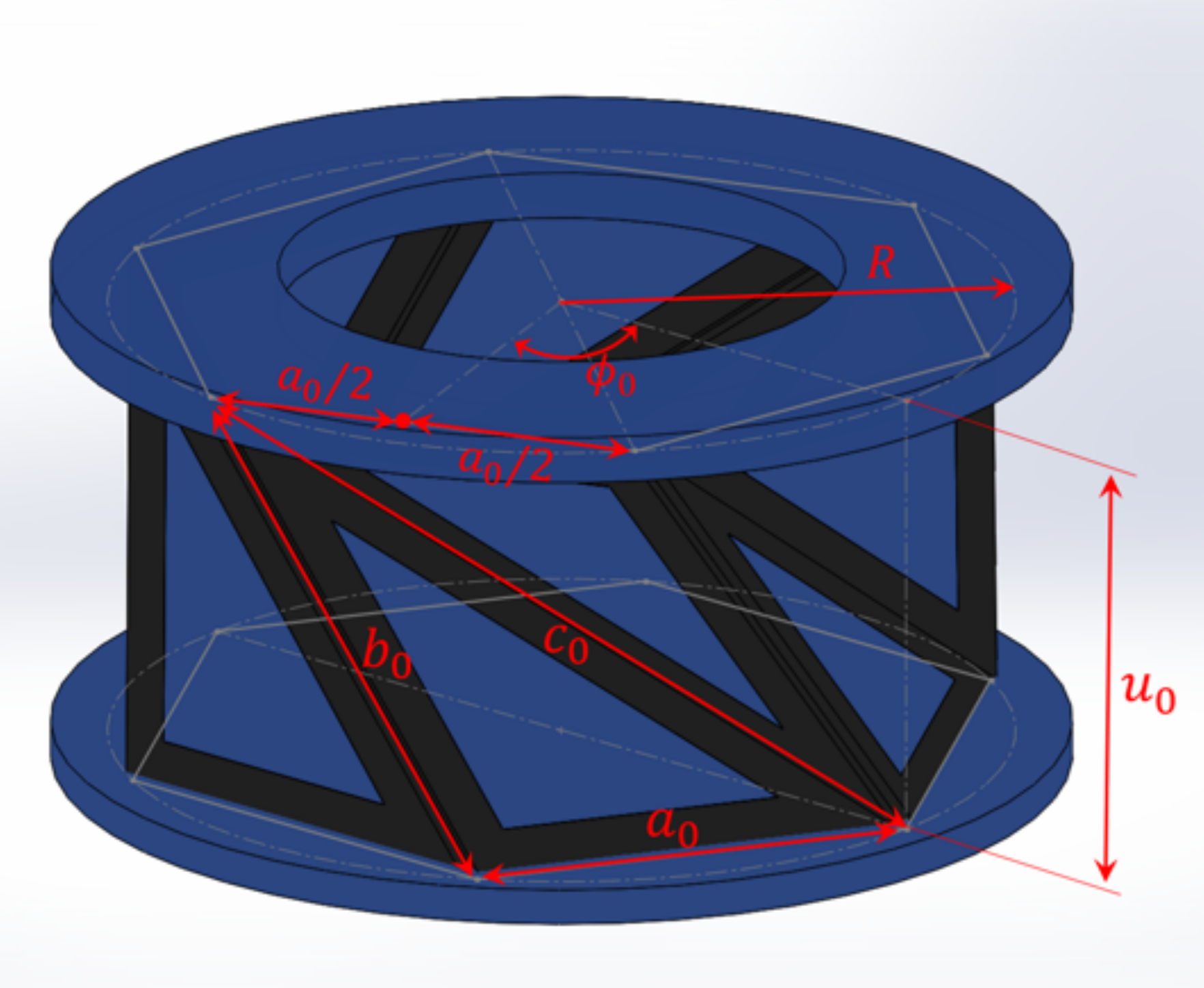}
		\caption{The fundamental geometric design parameters of the KOS.} 
		\label{Fund_Triangle}
	\end{figure*}

As shown in Fig. \ref{Fund_Triangle}, the final design of the KOS is characterized by several geometric design parameters. These are $i)$ the number of sides, $n$, of the parallel polygons; $ii)$ the radius, $R$, of the circle circumscribing those polygons; $iii)$ the vertical distance, $u_0$, and relative angle, $\phi_0$, between the end planes; and, finally, $iv)$ the length of the three edges, $a_0$, $b_0$, and $c_0$, forming the fundamental triangles that together form the KOS. Throughout this manuscript, the distance $u_0$ and the angle $\phi_0$ will be loosely referred to as the design height and design angle, respectively.

Depending on the values of the aforementioned geometric design parameters, a KOS can exhibit qualitatively different restoring force behavior. For instance, certain combinations of the geometric parameters can yield a single equilibirum configuration; while others may involve the presence of multiple equilibrium configurations. In either case, it has been shown that the number of the equilibrium configurations of the KOS can be well-approximated by adopting an axial truss model wherein each triangle in the KOS is represented by axially-deformable truss elements placed at its edges \cite{YasudaKresling2017,RmasanaKIOS19}. \footnote{The reader shoud keep in mind that the axial truss model was used in this paper only for its ability to provide a qualitative understanding of the general behavior of KOSs, and as a guideline for the choice of the design parameters that lead to qualitatively different behaviors. For the many reasons mentioned in Ref.  \cite{RmasanaKIOS19}, the linear truss model cannot be used as a means for quantitative prediction of the KOS behavior.} Using the truss model, the relative position and orientation of the two end planes during deployment can be fully characterized by describing the length of the three edges of the triangle, $a$, $b$, and $c$, in terms of the other design parameters as,
\begin{equation}
\begin{split}
a&=2R\sin{\frac{\pi}n},\\
b&=\sqrt{4R^2\sin^2\left({\frac{\phi-\frac{\pi}n}2}\right)+u^2},\\
c&=\sqrt{4R^2\sin^2\left({\frac{\phi+\frac{\pi}n}2}\right)+u^2}. 
\label{edge_lengths}
\end{split}
\end{equation}
where $\phi$ and $u$ are, respectively, the relative angle and vertical distance between the end planes under loading. 

Assuming that the base of each triangle does not undergo any deformation during deployment, and that the panels do not buckle under compressive loads, nor do they experience self avoidance at small values of $u$, the total strain energy stored due to panel deformation can be approximated by

\begin{equation}
    \Pi=\frac{nEA}2\left[\frac{(b-b_0)^2}{b_0}+\frac{(c-c_0)^2}{c_0}\right],
\end{equation}

where $EA$ is the axial rigidity of the truss elements.

\begin{figure*}[htb!]
		\centering 
		\includegraphics[width=12.0cm]{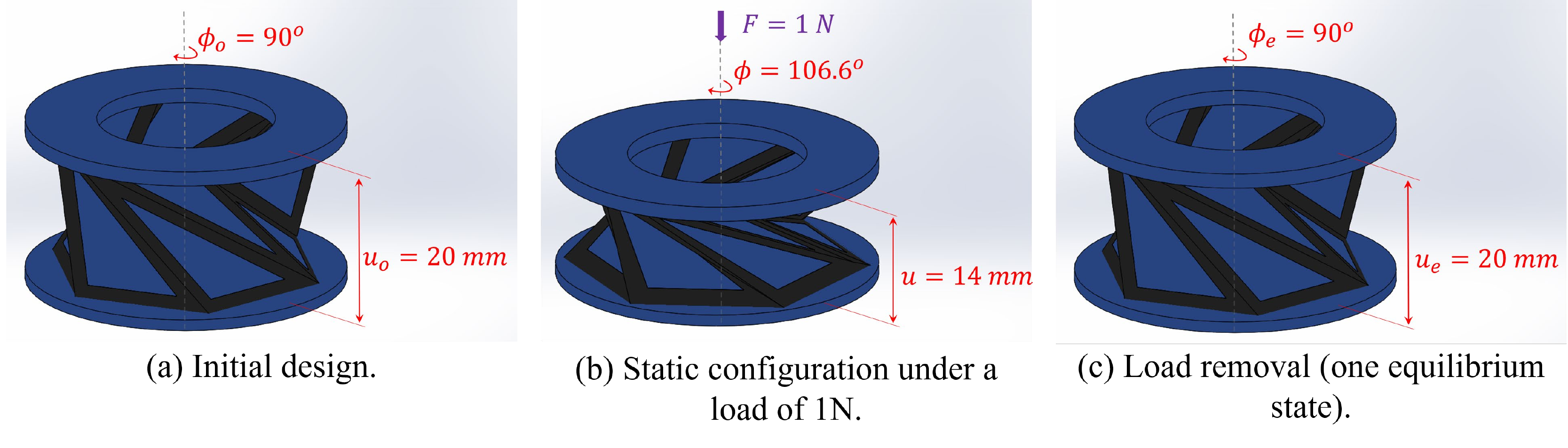}
		\caption{Schematic of a mono-stable KOS designed using the parameters $\phi_0=90^\circ$, $u_0=20$ mm, and $n=6$.}
		\label{mono}
	\end{figure*}

\begin{figure*}[htb!]
		\centering 
		\includegraphics[width=12.0cm]{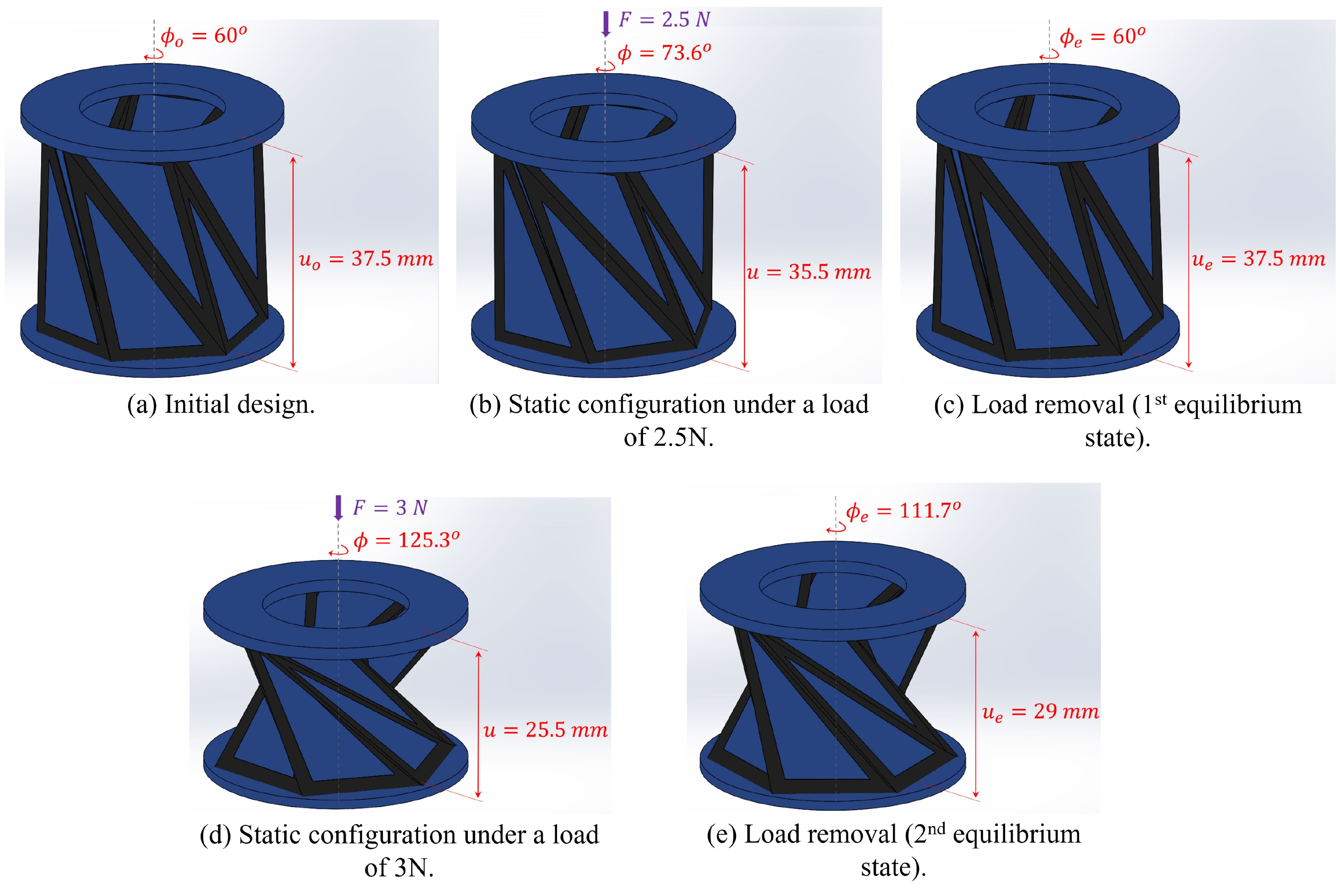}
		\caption{Schematic of a bi-stable KOS designed using the parameters $\phi_0=60^\circ$, $u_0=37.5$ mm, and $n=6$.}
		\label{bi}
	\end{figure*}

For given geometric design parameters, the possible stationary configurations or equilibrium states ($u_e, \phi_e$) of the KOS occur at the points which minimize the strain energy; i.e. $ \Pi_u|{(u_e,\phi_e)}= \Pi_\phi|{(u_e,\phi_e)}=0$. A possible stationary configuration is physically realizable (stable) if and only if it represents a minimum in the strain energy; i.e. $ \Pi_{uu}  \Pi_{\phi \phi}|{(u_e,\phi_e)}- \Pi_{u\phi}^2|{(u_e,\phi_e)}>0$ and $ \Pi_{uu}>0$. Based on this understanding, KOSs can be classified into the following two major categories: 
\begin{itemize}
\item \underline{Mono-stable KOS:} Such KOS has only one stable equilibrium state. This implies that, in the absence of external loads, the KOS can only settle at a single equilibrium position around which it can be operated. Figure~\ref{mono} depicts the schematic of a mono-stable KOS with the design parameters $\phi_0=90^\circ$, $u_0=20$ mm, and $n=6$. When a compressive load of 1 N is applied to the top polygon, the KOS undergoes simultaneous rotation and deflection until the restoring force balances the applied load at $\phi=106.6^\circ$ and $u=14$ mm. Upon removal of the load, the KOS springs back to the initial configuration which represents its only equilibrium configuration. 
 
\item  \underline{Bi-stable KOS:} Such KOS has two stable equilibrium states. This implies that, in the absence of external loads, the KOS can settle at either one of the two equilibrium points. The KOS can be operated around either one of those states or can be forced to jump between them. Figure~\ref{bi}(a) depicts the schematic of a bi-stable KOS designed using $\phi_0=60^\circ$, $u_0=37.5$ mm, and $n=6$. When a compressive load of 2.5 N is applied to the top polygon, the KOS undergoes simultaneous rotation and deflection until the restoring force balances the applied load at $\phi=73.6^\circ$ and $u=35.5$ mm. Upon removal of the load, the KOS springs back to its initial configuration which represents one of its equilibrium configurations. The second equilibrium  configuration can be achieved by applying a larger static compressive load; for instance, 3~N and removing it as shown in Fig.~\ref{bi} (e). This causes the KOS to settle at a different equilibrium state corresponding to $(\phi_e=111.7^\circ, u_e=29$ mm).
\end{itemize}

\begin{figure*}[htb!]
		\centering 
		\includegraphics[width=12cm]{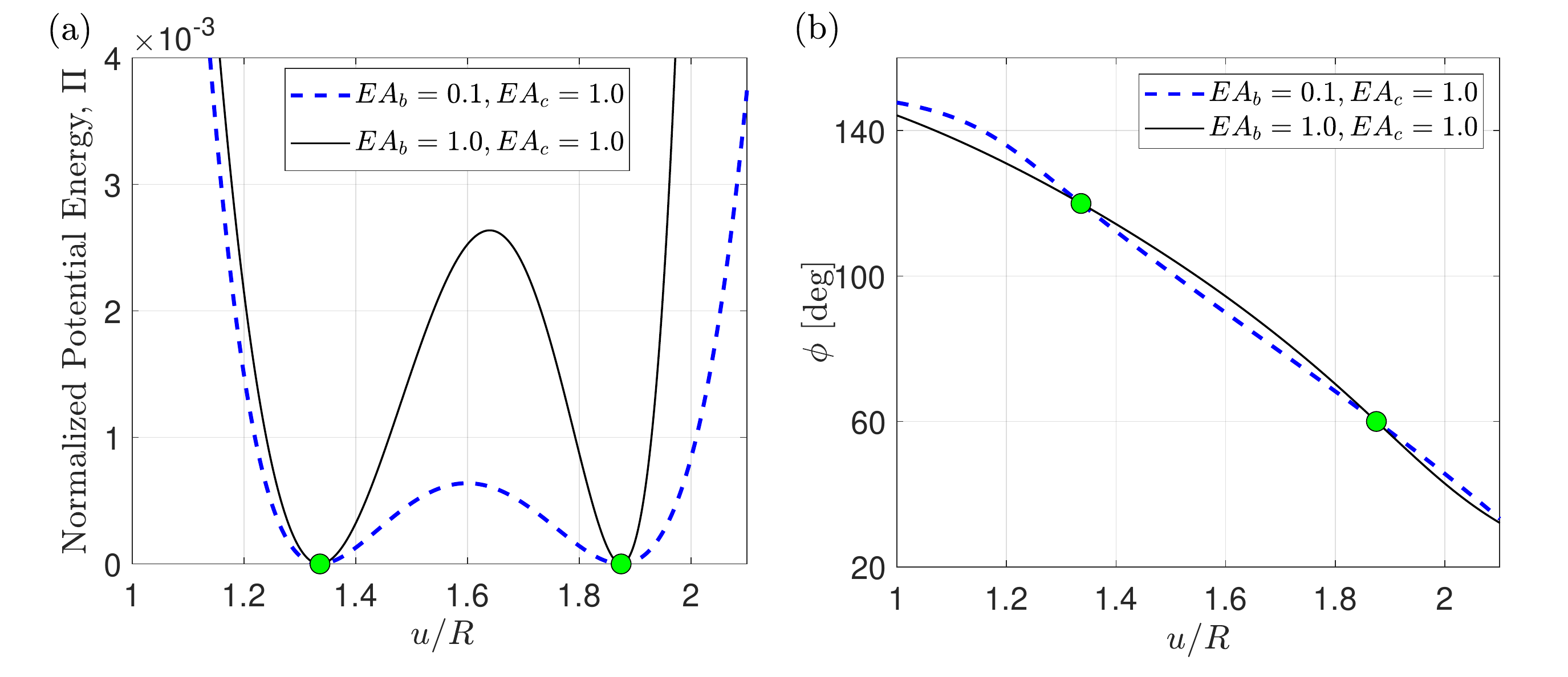}
		\caption{Effect of the axial rigidity on (a) the potential energy function and (b) path of rotation of a bi-stable KOS designed using the parameters $\phi_0=60^\circ$, $u_0/R=1.875$ mm, and $n=6$.} 
		\label{Kirigami}
	\end{figure*}

It turns out that the axial truss model yields equilibrium states that are invariant under changes in the material properties. That is, regardless of how the axial rigidity of the trusses is changed, the static equilibria remain the same. Note that this is different from saying that the quasi-static behavior of the KOS does not change with the material properties. In fact, the behavior of the KOS away from the equilibrium states is very sensitive to the axial rigidity.  Figure~\ref{Kirigami} clearly demonstrates this fact by showing that, while the shape of the potential energy function and associated restoring force change considerably with the axial rigidity, the equilibrium points remain the same.

\section{Design Map}	
\label{maps}
\begin{figure*}[htb!]
		\centering 
		\includegraphics[width=12cm]{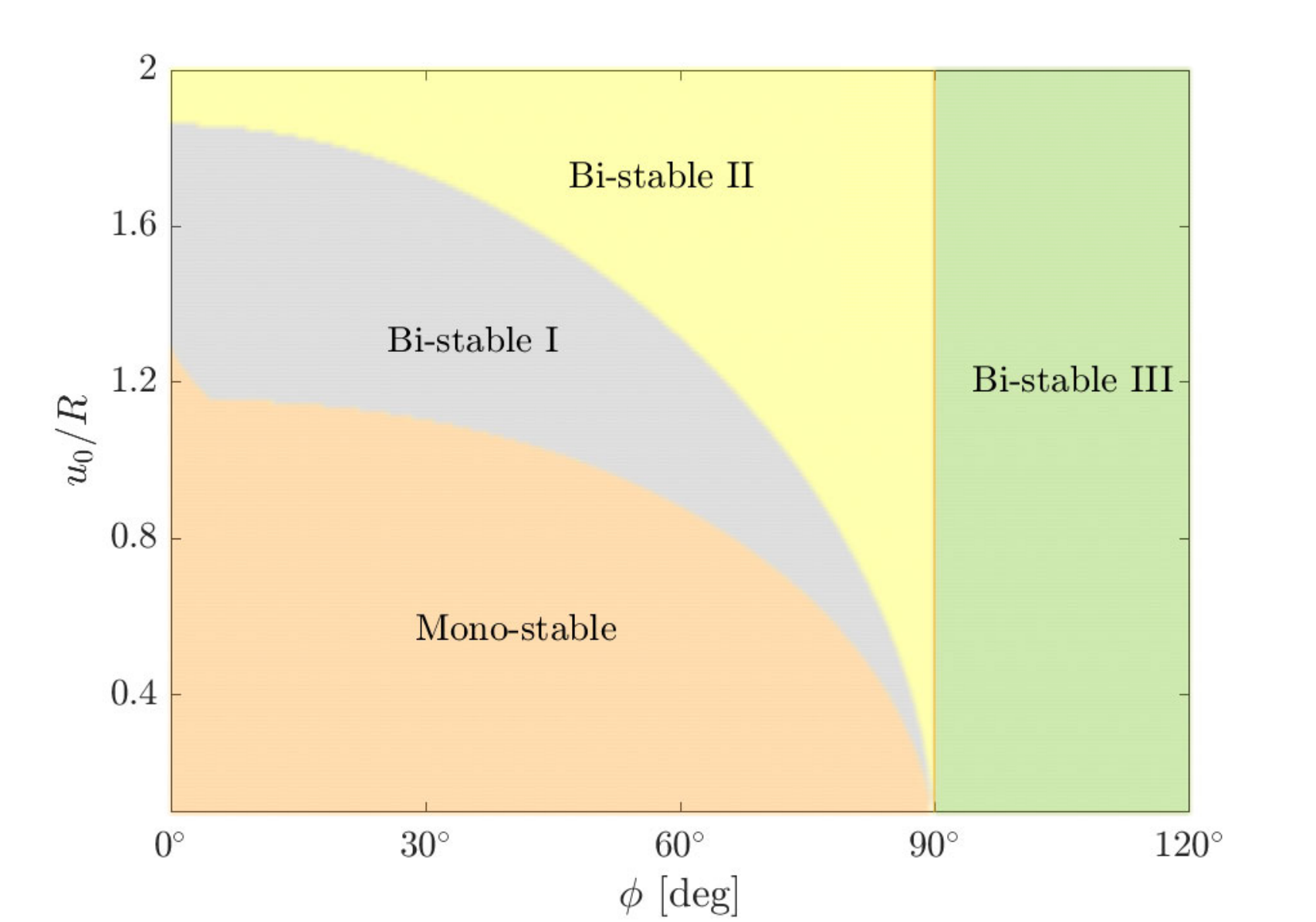}
		\caption{A design map demarcating regions in the design parameters space which lead to KOSs of  different potential characteristics for $n=6$.} 
		\label{Designmap}
	\end{figure*}

Using the axial truss model, it is possible to create design maps that demarcate the design space ($u_0/R, \phi_0$) of the KOS into different regions based on whether the KOS is mono- or bi-stable. Figure \ref{Designmap} depicts one such map for $n=6$. The map illustrates that, when the geometric design of the spring is such that, $\phi_0 <90^\circ$, the KOS can either be mono- or bi-stable depending on the value of, $u_0/R$. For relatively small values of $u_0/R$, the KOS is mono-stable but becomes bi-stable as $u_0/R$ increases. For values of $\phi_0$ greater than $90^\circ$, the KOS is always bi-stable regardless of the value of $u_0/R$. At $\phi_0=90^\circ$, the two equilibrium points associated with the bi-stable KOS coalesce and it becomes mono-stable.

\begin{figure*}[htb!]
		\centering 
		\includegraphics[width=12.0cm]{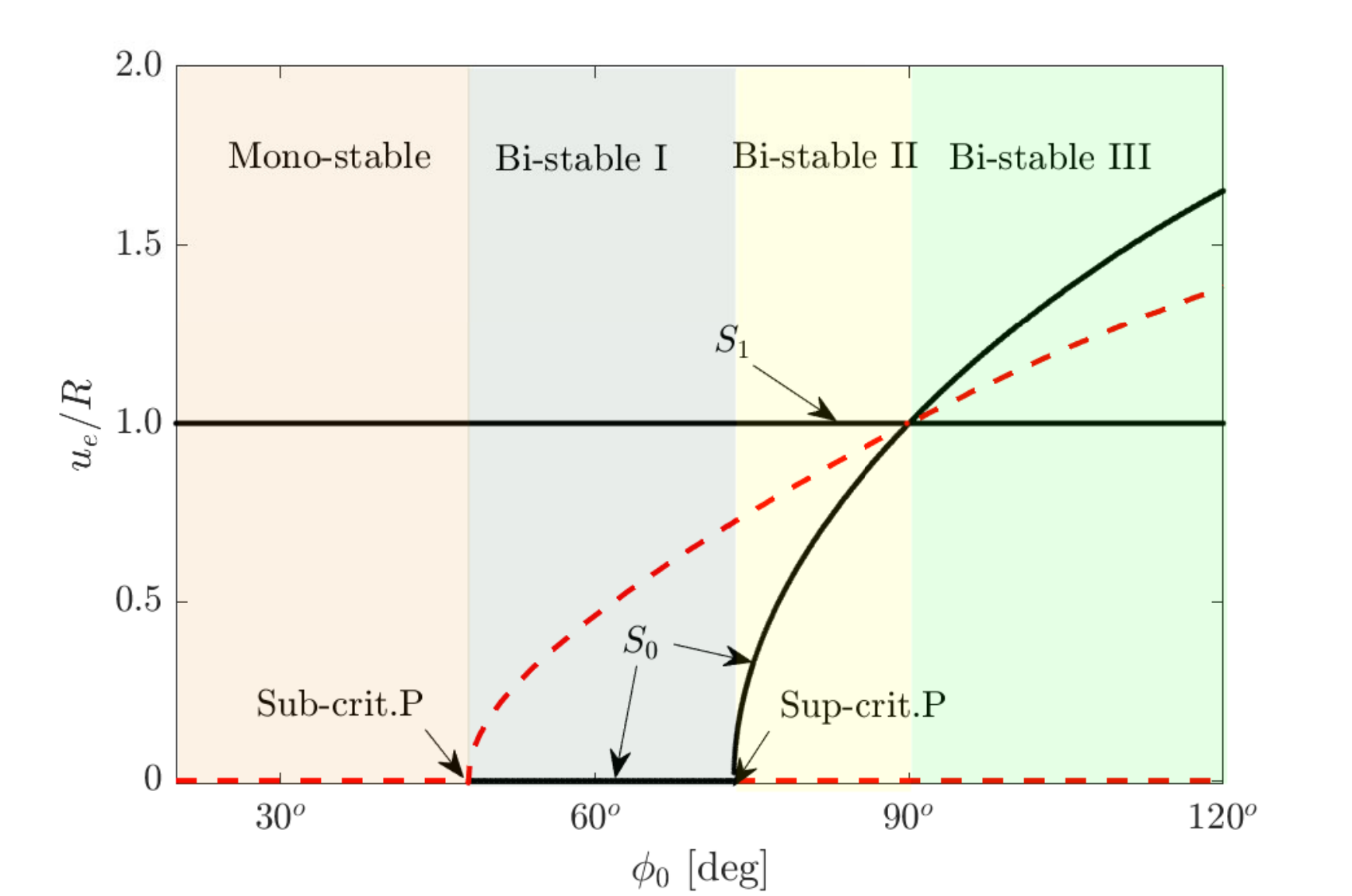}
		\caption{Variation of the KOS equilibria with the design angle, $\phi_0$. Results are obtained for a KOS with a design height, $u_0/R=1.0$.}
		\label{Bifurcation1}
	\end{figure*}

To further understand the design map shown in Fig. \ref{Designmap}, we study in Fig.~\ref{Bifurcation1} variation of the KOS equilibrium height, $u_e/R$,  as the design angle, $\phi_0$ is varied. The KOS is constructed such that $u_0/R=1$ and $\phi_0=20^\circ$. The design angle, $\phi_0$, is then increased to assess whether any new equilibrium points are born at other values of $\phi_0$. It can be clearly seen that $u_e/R=1$ is the only equilibrium height up to $\phi_0\approx 48^\circ$, where a new equilibrium point is born at $u_e/R=0$ due to a sub-critical pitchfork (Sub-crit.P) bifurcation. As a result, the KOS becomes bi-stable with two different equilibrium states: $u_e/R=0$ denoted here as $S_0$, and $u_e/R=1$ denoted here as $S_1$. This behavior continues until around $\phi_0=73^\circ$, where the lower trivial equilibrium state $S_0$ loses stability and is replaced by another nonzero equilibrium state $S_0>0$. The KOS remains bi-stable as $\phi_0$ is increased up to $\phi_0=90^\circ$. At this point, $S_0$ and $S_1$ coalesce  and the KOS becomes mono-stable. Further increase in $\phi_0$ causes $S_0$ to become larger than $S_1$. 

Depending on the relative magnitude of $S_0$ and $S_1$, the bi-stable region can be classified into three distinct areas: bi-stable I wherein $S_0$ is always zero and $S_1$ is nonzero, bi-stable II wherein $S_0$ and $S_1$ are both non-zero, but $S_0<S_1$, and  bi-stable III  wherein $S_0$ and $S_1$ are both non-zero, but $S_0>S_1$. 
  
\section{Fabrication}
\label{fabrication}
A KOS with the desired quasi-static behavior can be fabricated by carefully selecting the geometric parameters from the map shown in Fig. \ref{Designmap}. This can be achieved using any flat sheet of material which can be creased and that can stretch or compress along the creases while retaining the creases \cite{RmasanaKIOS19}. Paper, that is traditionally used for origami, serves this purpose. However, because of its low rupture strain of about $2\%$, it develops permanent kinks when the strain limits are exceeded. As a result, after a few finite number of loading and unloading cycles, the KOS ruptures at the folds.  This, combined with the low load bearing capabilities of paper, and other foldable materials, preclude their usage in actual engineering applications. Similarly, while traditional engineering materials like sheet metals could provide for the load bearing capabilities, they tend to be very stiff, hard to fold, and have low fatigue life for the kind of stresses that a KOS undergoes especially at the folds. Furthermore, while traditional origami principles based on folding and creasing can be used to build structures for demonstrative purposes, fabrication errors and uncertainties among the samples are unavoidable because of human involvement \cite{RmasanaKIOS19}. To avoid those issues and uncertainties, we employed 3-D printing in conjunction with a special design to fabricate a KOS which mimics the qualitative behavior of the paper-based folded springs without compromising on functionality or durability. To the author's knowledge, the proposed design represents the first manufacturable, durable, and fully-functional origami-inspired KOS.

To achieve this goal, each of the fundamental triangles is redesigned to accommodate for easy folding and stretching at the panel interfaces while providing sufficient rigidity such that it follows the Kresling origami functionality without collapsing under loading. Fabrication was performed using a Stratasys J750 3-D printer implementing the polyjet technique, which involves depositing extremely thin layers of photopolymer and exposing each layer to UV light for immediate curing and hardening \cite{singh2011process}. With this technique, the 3-D printer is capable of producing structures with complex shapes, smooth surfaces, and detailed features. Structures of sizes as large as 490 $\times$ 390 $\times$ 200 mm can be built, and layer thickness can be as low as 14 microns \cite{stratasys}. Moreover, different parts of the structure can be printed with either flexible or rigid materials, or a mixture of both for tuning the amount of flexibility/rigidity needed, making it the ideal 3-D printing technology to serve our purpose. Using this technology, each panel was fabricated using two materials: $i)$ an inner central rigid core which was fabricated using \textit{Vero}, a rigid plastic polyjet material; and $ii)$ an outer frame which was fabricated using \textit{TangoBlackPlus}, a flexible rubber-like polyjet material. The outer frame is made from the soft material so as to permit folding and stretching at the interfaces.

\begin{figure*}[htb!]
		\centering 
		\includegraphics[width=12.0cm]{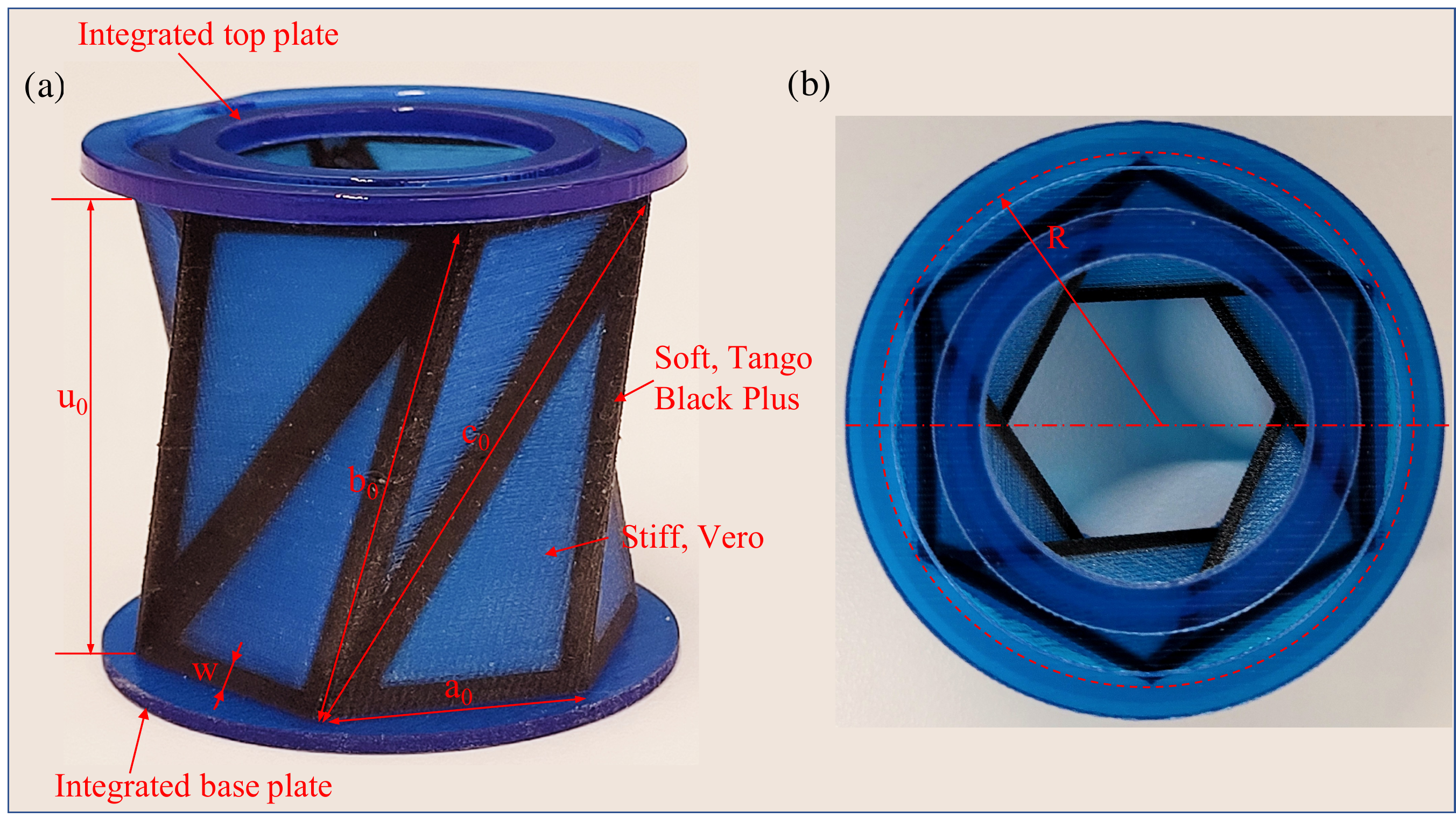}
		\caption{A 3-D fabricated KOS. (a) Side view, and (b) top view.}
		\label{3D}
	\end{figure*}

After the 3-D printing of the KOS is completed, the support material is then removed by waterjetting, and the structure is placed in a UV light chamber for 300 minutes to ensure that the materials are fully cured.  An example of the fabricated KOS is shown in Fig. \ref{3D}. It can be clearly seen that the fabricated KOS looks exactly similar to the one fabricated by paper folding. Nevertheless, the new design contains additional geometric design parameters; namely, the width, $w$, and thickness $t$,  of the soft edges. In what follows, we study the quasi-static behavior of the fabricated KOSs for different geometric design parameters and highlight the influence of $w$ and $t$ on the quasi-static behavior.

\section{Quasi-static Behavior}
\label{quasi}
\begin{figure*}[htb!]
		\centering 
		\includegraphics[width=6.0cm]{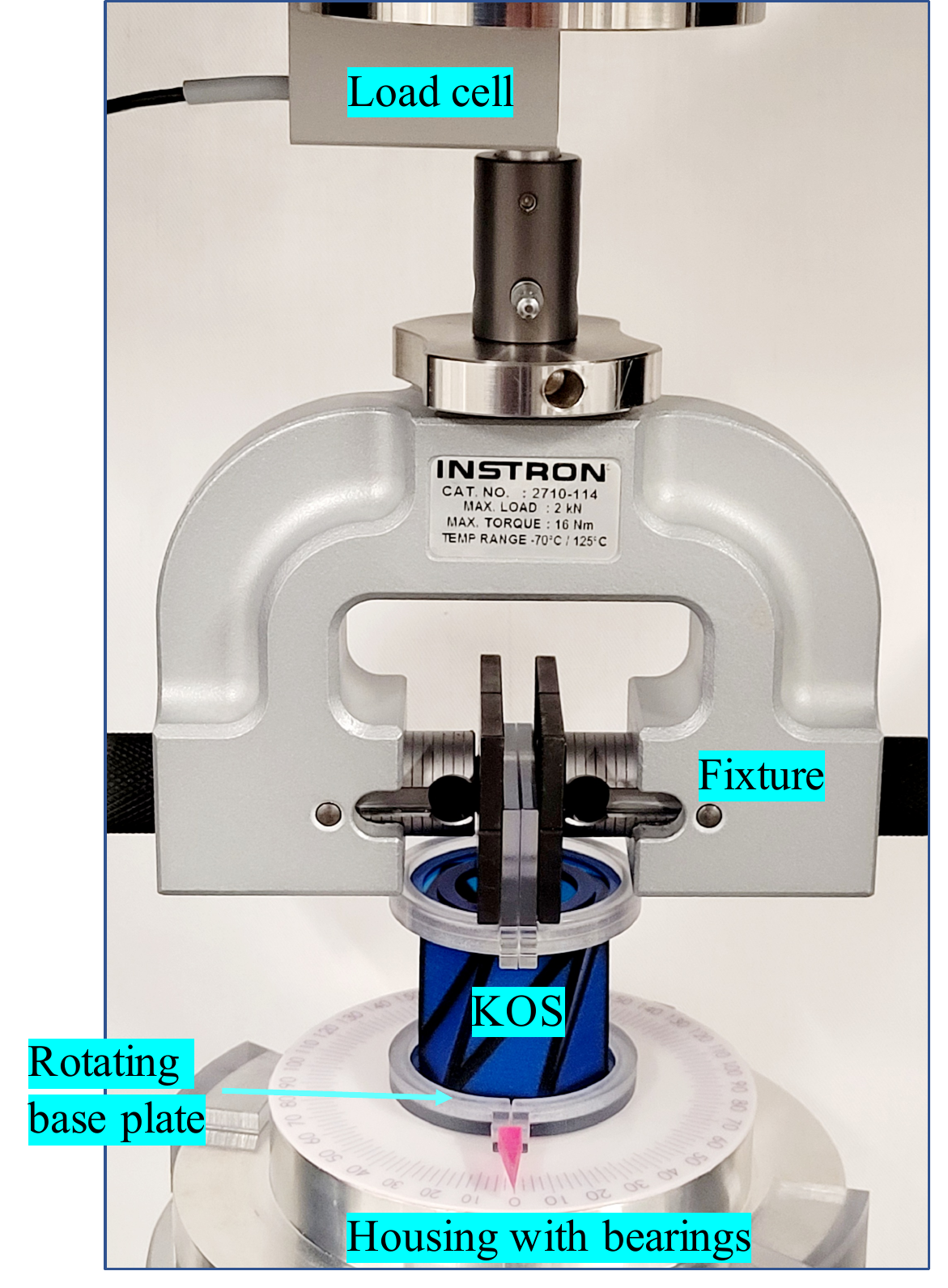}
		\caption{Experimental setup used for quasi-static testing.}
		\label{ExpSetup}
	\end{figure*}

The restoring force behavior of the KOS is determined experimentally via compressive and tensile tests performed on a universal testing machine. During the tests, the bottom end is placed on a platform that is free to rotate about a centroidal axis, while the top end of the KOS is subjected to a controlled fixed rate displacement of $0.2$ mm/s. The restoring force is then measured using a load cell as shown in Fig.~\ref{ExpSetup}.

During the experiments, the zero displacement reference is chosen to be the initial undeformed position ($u_0,\phi_0$) when the applied load is zero. It is observed that the restoring force of the KOS is slightly different during compression and tension cycles. Hence, the load is applied in a cyclic manner and the responses in compression and tension are both recorded.  Positive force measurements on the restoring force curves represent compressive loads while negative measurements represent tensile loads. A set of five 3-D printed samples are tested for the same design parameters. The average and standard deviation of the data are presented for each design. In the following subsections, we present the experimentally identified restoring force curves for the qualitatively different KOS behavior discussed in the design map shown in Fig.~\ref{Designmap}.

\subsection{Mono-stable KOS}

	\begin{figure*}[htb!]
		\centering 
		\includegraphics[width=12cm]{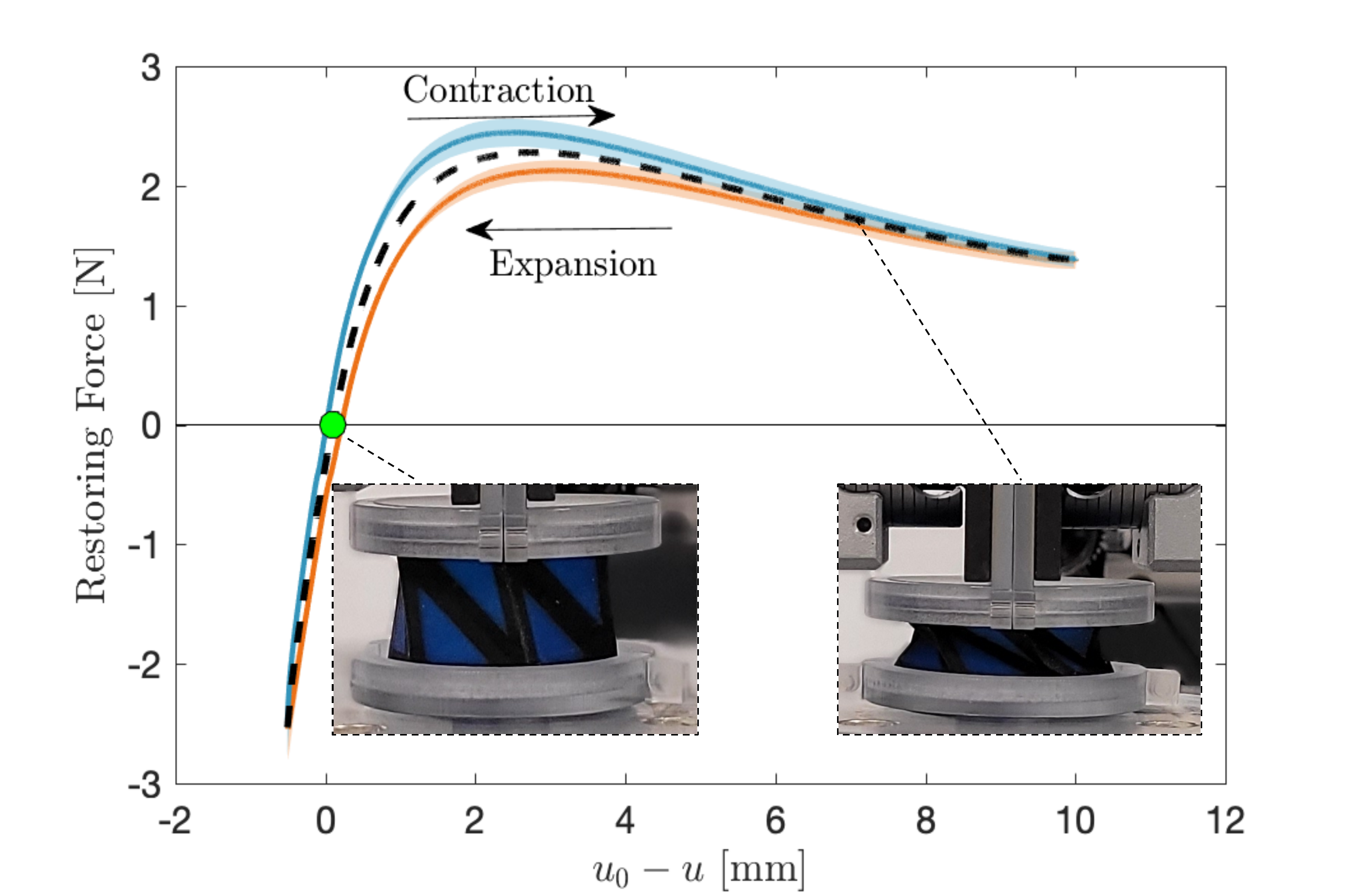}
		\caption{Restoring force curve of a mono-stable KOS designed using the parameters, $u_0/R=1.0$, $\phi_0=45^o$,  $n=6$, $R=20$ mm, $t=1$ mm and $w=2$ mm. Insets show the 3-D printed KOS at two different deployment heights.}
		\label{M-type}
	\end{figure*} 	
	
A mono-stable KOS can be fabricated using any set of geometric parameters selected from the mono-stable region of Fig.~\ref{Designmap}. Here, we make an arbitrary choice of $u_0/R=1.0$, $\phi_0=45^o$, $n=6$, and $R=20$ mm. It should be noted that, unless otherwise specified, the number of sides of the KOSs used in all subsequent experiments is $n=6$ and the radius of circumscribing circle is $R=20$ mm. Figure~\ref{M-type} depicts the measured restoring force curve along with the average and standard deviation of the five samples used. The first thing observed when inspecting the curves is that the standard deviation of the data among the different samples is extremely small, which points to the repeatability of the fabrication process. For this mono-stable KOS, the initial undeformed position is the only equilibrium state.  The restoring force around this equilibrium is nearly linear for small deflections and the stiffness is approximately $0.6$ N/mm. At around a deflection of $0.5$ mm ($u_0-u=2.5\% u_0$), the response begins to soften in compression and the measured force starts to decrease. In tension, any deflections beyond $0.5$ mm causes excessive stretching that would ultimately damage the KOS. As such, for this combination of design parameters, the KOS is better suited to carry compressive loads and would behave linearly for up to $2.5\% u_0$. 

\begin{figure*}[htb!]
		\centering 
		\includegraphics[width=12.0cm]{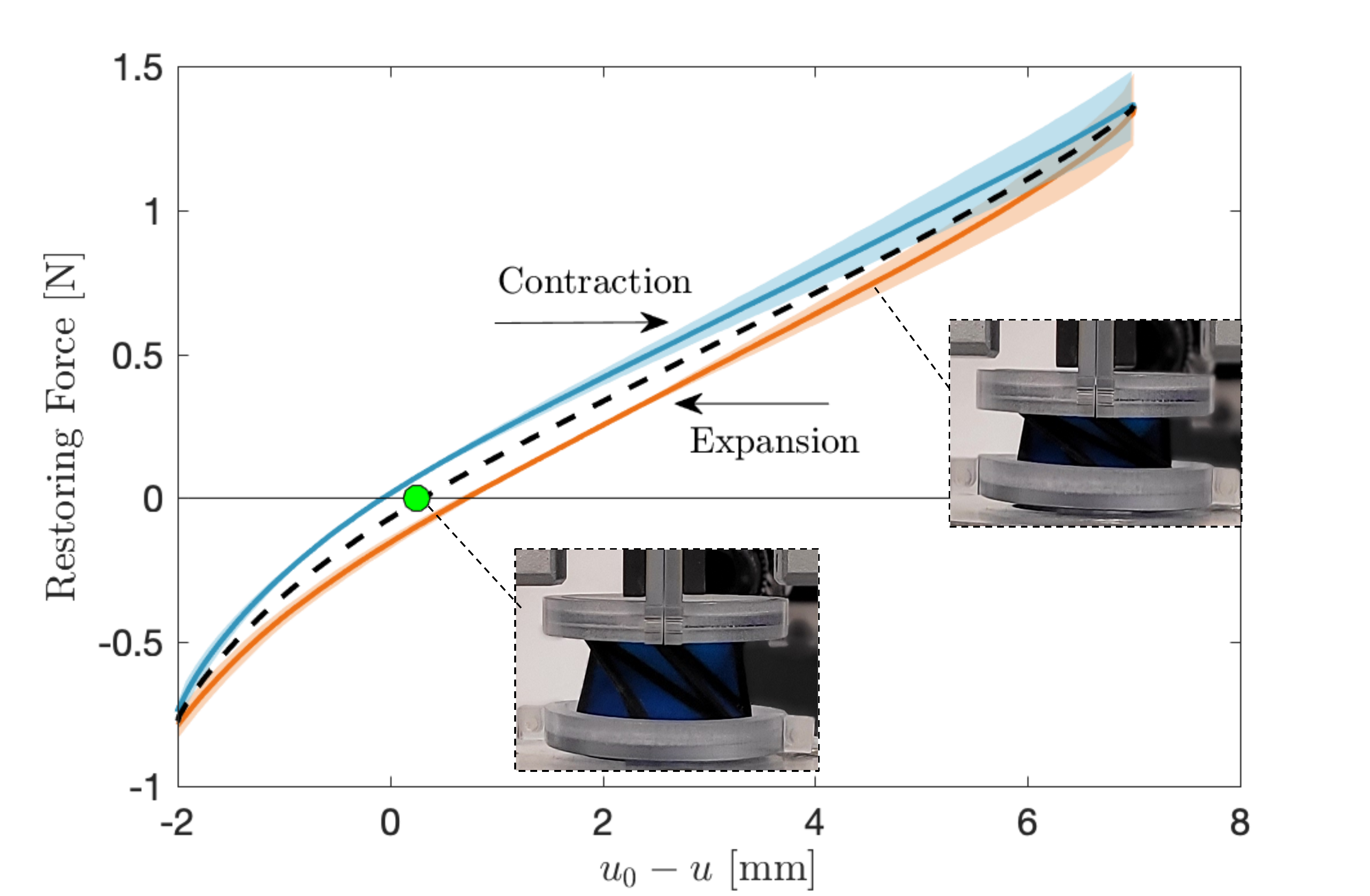}
		\caption{Restoring force curve of a mono-stable KOS designed using the parameters, $u_0/R=1.0$, $\phi_0=90^o$,  $n=6$, $R=20$ mm, $t=1$ mm and $w=2$ mm. Insets show the 3-D printed KOS at two different deployment heights.} 
		\label{M1}
	\end{figure*}

Another mono-stable KOS can be constructed using $\phi_0=90^\circ$ and any value of $u_0$. Theoretically, choosing these design parameters would result in a mono-stable KOS with nearly zero stiffness. While  the experimental results shown in Fig.~\ref{M1} do not demonstrate the zero stiffness behavior predicted theoretically; still the resulting linear stiffness of $0.2$ N/mm is small when compared to that associated with Fig.~\ref{M-type}.  Compared to the mono-stable KOS whose restoring force behavior is shown Fig.~\ref{M-type}, this spring has a relatively much larger range of linear behavior with lower stiffness. Such behavior is ideal for applications requiring low stiffness linear behavior over a wide displacement range. Examples of such applications include, but are not limit to, pistons used in dispensing medicines \cite{ROSE1999411}, and seismographs \cite{LorainSeismograph}.

\subsection{Bi-stable KOS}
\begin{figure*}[htb!]
		\centering 
		\includegraphics[width=12.0cm]{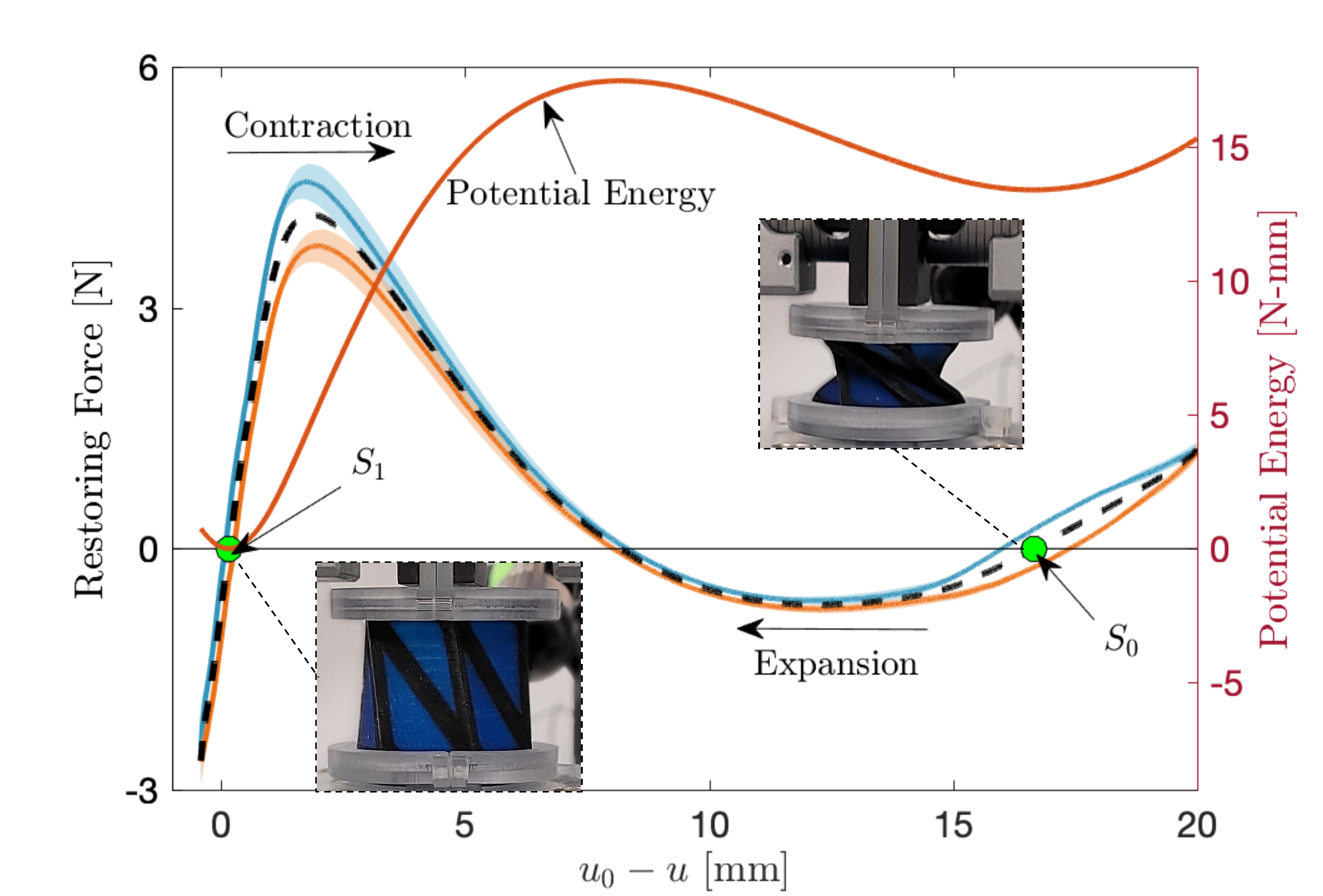}
		\caption{Restoring force curve and calculated potential energy of a bi-stable KOS designed using the parameters, $u_0/R=1.65$, $\phi_0=45^o$,  $n=6$, $R=20$ mm, $t=1$ mm and $w=2$ mm. Insets show the 3-D printed KOS at two different equilibria.} 
		\label{B-type}
	\end{figure*}

A bi-stable KOS can be fabricated using any set of geometric parameters selected from the bi-stable region of the map shown in Fig.~\ref{Designmap}. First, we fabricated a KOS with the design parameters, $u_0/R=1.65$, $\phi_0=45^o$, $R=20$ mm, and $n=6$, which fall in the bi-stable I region. Following the experimental procedure described earlier, the restoring force curve of the bi-stable KOS is generated as shown in Fig.~\ref{B-type}. The figure demonstrates that the spring has two stable equilibrium points $S_0$ and $S_1$, where $S_1$ occurs at the design height, $u_0$, while $S_0$ occurs at a deflection $u_0-u\approx 16.5$ mm. These two equilibria represent local minima in the strain energy of the KOS.

Around $S_1$, the spring is very stiff ($4.5$ N/mm) with a nearly linear behavior.  Upon compressing the KOS beyond $2$ mm, the restoring force saturates and starts to decrease causing a strong softening behavior that continues up to a deflection of $8$ mm at which the restoring force becomes zero. Note that this point represents an unstable equilibrium state of the KOS at which the strain energy is a local maximum. Beyond this point, the restoring force becomes negative implying that the KOS is being pulled towards the other stable equilibrium, $S_0$. Operating the spring around $S_0$ yields a nearly linear low-stiffness response ($0.4$ N/mm). Upon reversing the direction of the prescribed displacement, the KOS's quasi-static response tracks back the restoring force curve with slight offset due to the hysteresis in the viscoelastic material.

Any small quasi-static or dynamic disturbances applied to the KOS when it is settled at either $S_0$ or $S_1$ would result in small motions restricted to the potential energy well of the corresponding equilibrium point. However, if such motions are large enough to trigger escape from the potential well, the KOS moves from state $S_1$ to $S_0$ or vice versa. For this particular geometry, escape from state $S_0$ is easier as it requires less energy. Such a behavior can be used in applications related to sensitive mechanical switches for memory boards \cite{YasudaKresling2017,MasanaKIMS}, peristaltic robots \cite{BHOVAD2019100552}, robots for medical surgeries \cite{Bistable_medical1}, and energy harvesting devices \cite{DaqaqNVEHReview}.

	\begin{figure*}[htb!]
		\centering 
		\includegraphics[width=12.0cm]{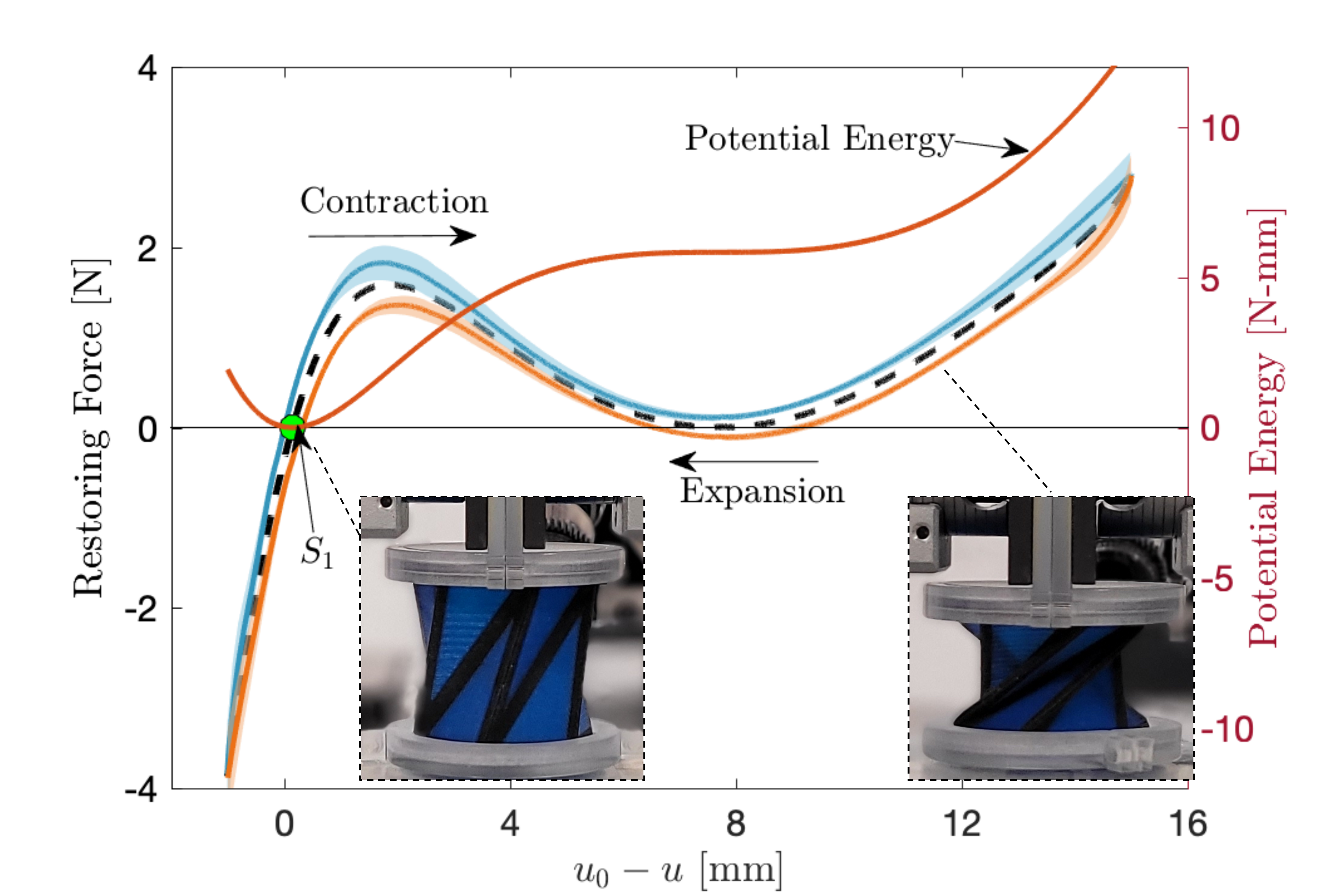}
		\caption{Restoring force curve and calculated potential energy of a bi-stable KOS designed using the parameters, $u_0/R=1.875$, $\phi_0=60^o$,  $n=6$, $R=20$ mm, $t=1$ mm and $w=2$ mm. Insets show the 3-D printed KOS at two different deployment heights.} 
		\label{B1}
	\end{figure*}

Another KOS is fabricated with its design parameters selected from the bi-stable II region of the map shown in Fig. \ref{Designmap}, using $u_0/R=1.875$, and $\phi_0=60^\circ$. According to the design map generated using the simplified axial truss model, the KOS should behave as a bi-stable spring. However, upon fabricating the KOS with the said design parameters, it was found that the equilibrium, $S_0$, is only realizable in the expansion part of the displacement sweep as shown in Fig.~\ref{B1}. As such, the KOS is nearly but not exactly bi-stable. The potential energy function associated with this KOS is obtained by integrating the restoring force based on its average value during compression and tension. It is shown that the potential well associated with $S_0$ is extremely shallow, and, hence the KOS can be easily perturbed to escape $S_0$ and settle at $S_1$. The difference between the predictions of the theoretical model and the experiments is mainly due to the simplicity of the axial truss model, which does not account for residual stresses, buckling of the panels, and panel avoidance at small values of $u$. 

As aforementioned, when the design parameters of the KOS are taken from the bi-stable I region, one of the stable states, $S_0$, corresponds to $u=0$; while the other state $S_1$ corresponds to the design value, $u_0$. As such,  in theory, it is possible to maximize the deployable range of the bi-stable spring by choosing a large value of $u_0$ while choosing the initial angle $\phi_0$, such that the geometric design parameters fall in region I of the map. However, in practice, $u$ can never be exactly zero. This is because, as the two parallel end planes are pushed closer to each other, they compress and trap the panels in between. The self-avoidance of the panels, which is not accounted for in the simple truss model, prevents the KOS from achieving the theoretical zero value of $u$. 

\begin{figure*}[htb!]
		\centering 
		\includegraphics[width=12.0cm]{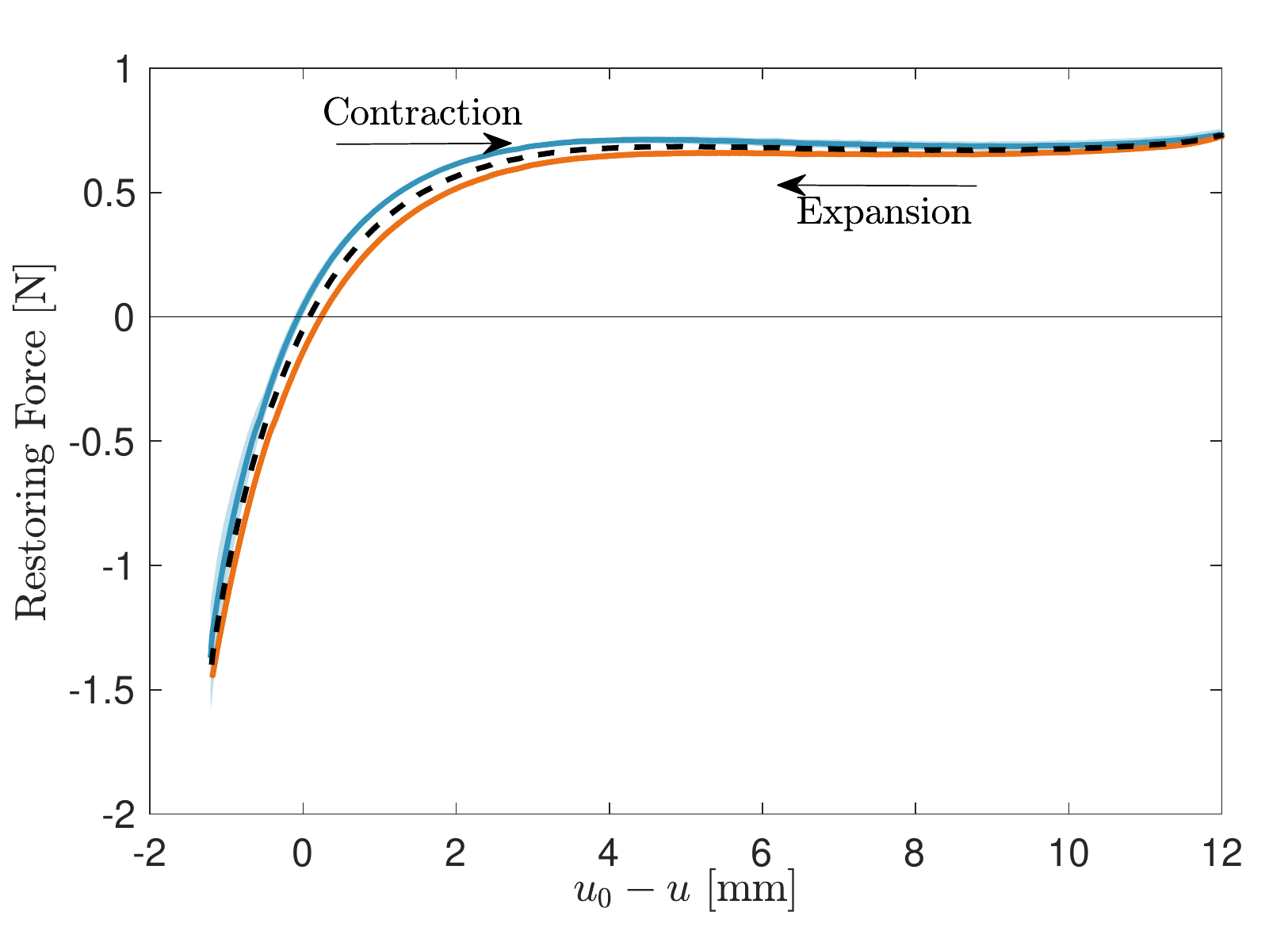}
		\caption{Restoring force curve of a KOS designed using the parameters: $u_0/R=1.1$, $\phi_0=65^o$,  $n=6$, $R=20$ mm, $t=1$ mm and $w=2$ mm.} 
		\label{MB}
	\end{figure*}

To study the effect of self avoidance on the restoring force curve, a KOS is fabricated using the geometrical parameters taken from the bi-stable I region, where $u_0/R=1.1$, and $\phi_0=65^\circ$. As shown in  Fig.~\ref{MB}, self avoidance of the panels prevents the formation of the equilibrium state $S_0$. Therefore, instead of having a bi-stable KOS, the KOS becomes mono-stable possessing a quasi zero stiffness (QZS) over a large range of deployment. This characteristic is particularly useful in designing effective vibration isolators \cite{QZSBook,CARRELLA2007678}, seismographs \cite{LorainSeismograph}, gravimeters \cite{LaCosteQZS}, and other statically balanced devices \cite{DorsserQZS,CarricatoQZS}. The zero stiffness is typically achieved over a small displacement range using a combination of positive and negative stiffness elements \cite{SchenkQZS}, which could be either in the form of oblique springs \cite{CARRELLA2007678}, or magnetic springs that provide the levitation to the sprung mass \cite{MagneticQZS}. In  \cite{ishida2017design}, the authors explored using Kresling origami inspired structures to develop QZS isolators; however, the various mechanical members including the joints used to construct the spring made the spring complex and vulnerable for design eccentricities. The 3-D printed KOSs may provide a viable approach to design QZS systems because of their simplistic yet high design feasibility.

\section{Effect of The Newly-Introduced Design Parameters}
\label{new}
\begin{figure*}[htb!]
		\centering 
		\includegraphics[width=12cm]{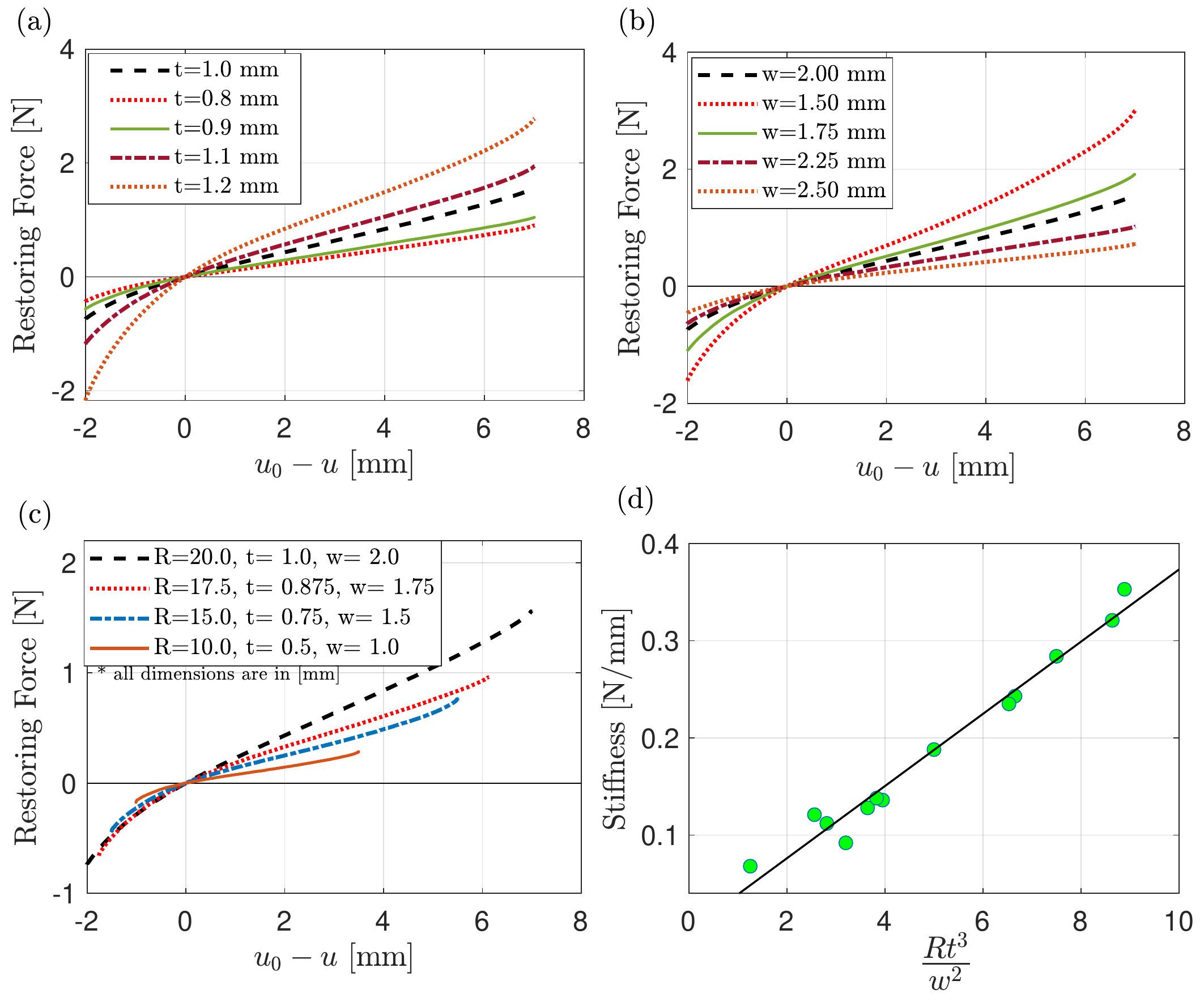}
		\caption{Variation of the restoring force with the displacement, $u_0-u$, for (a) different values of $t$, $w$ = $2$ mm, and $R$ = $20$ mm; (b) different values of $w$, $t$ = $1$ mm, and $R$ = $20$ mm; (c) $t$, $w$, and $R$ are scaled by the same amount; and (d) Dependence of the stiffness on $\frac{R t^3}{w^2}$. Results are obtained for a KOS designed using the parameters $\phi_0=90^\circ$, $u_0/R=1$, and $n=6$.   } 
		\label{M1_FabPar}
	\end{figure*}

To design the new KOSs, we introduced two additional design parameters; namely $w$, and $t$, which represent the thickness and width of the soft panel frame. These two parameters control the stiffness of the folds, and therewith the effective restoring force curve of the KOS. To understand the influence of these parameters on the quasi-static behavior of the mono- and bi-stable springs, we carried out a series of experiments on KOSs with different values of $w$ and $t$ while keeping the rest of the parameters constant. The first set was performed on a mono-stable KOS with the following design parameters: $\phi_0=90^\circ$, $u_0/R=1.0$, $R=20$ mm, and $w=2$ mm; while the thickness was changed around the nominal value of $t=1$. Figure~\ref{M1_FabPar}(a) demonstrates that the stiffness increases as the thickness of the panels is increased. The restoring force behavior is nearly linear for small values of $t$, but becomes nonlinear as $t$ is increased towards $t=1.2$~mm.

The second set of experiments was performed on the same KOS, but now the thickness $t$ and the radius $R$ were kept constant at $t=1$ mm and $R=20$ mm, respectively; while the width was changed around the nominal value of $w=2$ mm. Figure~\ref{M1_FabPar}(b) shows that the linear stiffness decreases as $w$ increases. The third set was performed on the same KOS with $\phi_0=90^\circ$, and $u_0/R=1.0$, but $t$, $w$, and $R$ were scaled together by the same amount. In other words, dimensionally-scaled versions of the KOS were created and tested. Figure \ref{M1_FabPar}(c) shows that the stiffness increases as the KOS is scaled up and decreases when it is scaled down. 

Based on the experimentally attained data, we observed that the linear stiffness of the KOS is directly proportional to the cube of the thickness; i.e. $k \propto t^3$, inversely proportional to the square of the width; i.e. $k \propto 1/w^2$, and directly proportional to $R$; i.e.  $k \propto R$. For the KOS designed using  $\phi_0=90^\circ$, $u_0/R=1.0$, and $n=6$, we found that

\begin{equation}
    k\approx \frac{1}{25} \frac{R t^3}{w^2},
\label{k-empirical}
\end{equation}

where the dimensions of $R$, $t$, and $w$ are measured in mm and $k$ is in N/mm. Figure~\ref{M1_FabPar}(d) shows the linear dependence of the stiffness on the parameter, $R t^3/w^2$.

\begin{figure*}[htb!]
		\centering 
		\includegraphics[width=12cm]{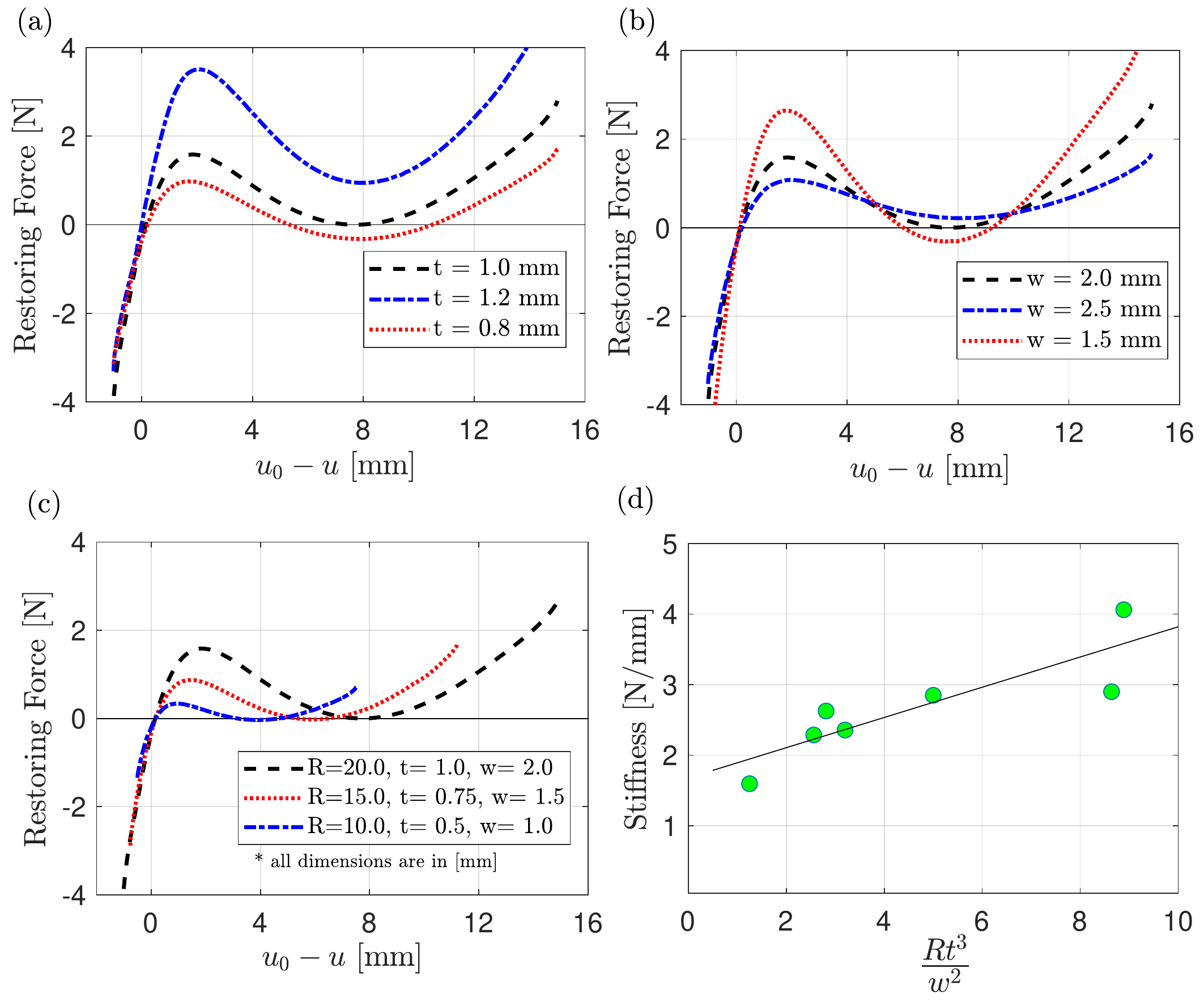}
		\caption{Variation of the restoring force with the displacement, $u_0-u$, for (a) different values of $t$, $w$ = $2$ mm, and $R$ = $20$ mm; (b) different values of $w$, $t$ = $1$ mm, and $R$ = $20$ mm; and (c) $t$, $w$, and $R$ are scaled by the same amount;  (d) Dependence of the stiffness on $\frac{R t^3}{w^2}$. Results are obtained for a KOS designed using the parameters $\phi_0=60^\circ$, $u_0/R=1.875$, and $n=6$. }
		\label{B1_FabriParam_Norm}
	\end{figure*}

In the case of the bi-stable KOS, we see from Fig.~\ref{B1_FabriParam_Norm} (a) that the stiffness around the equilibrium, $S_1$, does not change appreciably as the thickness is varied while keeping the rest of the parameters constant. On the other hand, the second equilibrium point, $S_0$, is considerably affected when changing, $t$. In specific, we see that the second equilibrium point ceases to exist when the thickness is increased from $t=0.8$ mm to $t=1.0$ mm. This stems from the fact that the potential wells become shallower and shallower as $t$ is increased up to the point where the KOS switches from the bi- to the mono-stable behavior. A similar behavior can be seen in Fig.~\ref{B1_FabriParam_Norm} (b) when decreasing the width while keeping the other design parameters constant. Quite interestingly, the qualitative behavior of the restoring force does not change when all the design parameters are scaled by the same amount as shown in  Fig.~\ref{B1_FabriParam_Norm} (c).  Scaling down the dimensions by the same amount only results in a shrinkage of the deployment range and a corresponding reduction in the magntitude of the restoring force. Again, Fig. ~\ref{B1_FabriParam_Norm} (d) shows the linear dependence of the stiffness on the parameter $R t^3/w^2$  around the equilibrium $S_1$.  It is found that for this bi-stable KOS,  the stiffness scales linearly with $R t^3/w^2$ as

\begin{equation}
    k\approx \frac{2}{9} \frac{R t^3}{w^2}.
\end{equation}

\section{Scaling}
\label{scaling}
\begin{figure*}[htb!]
		\centering 
		\includegraphics[width=12.0cm]{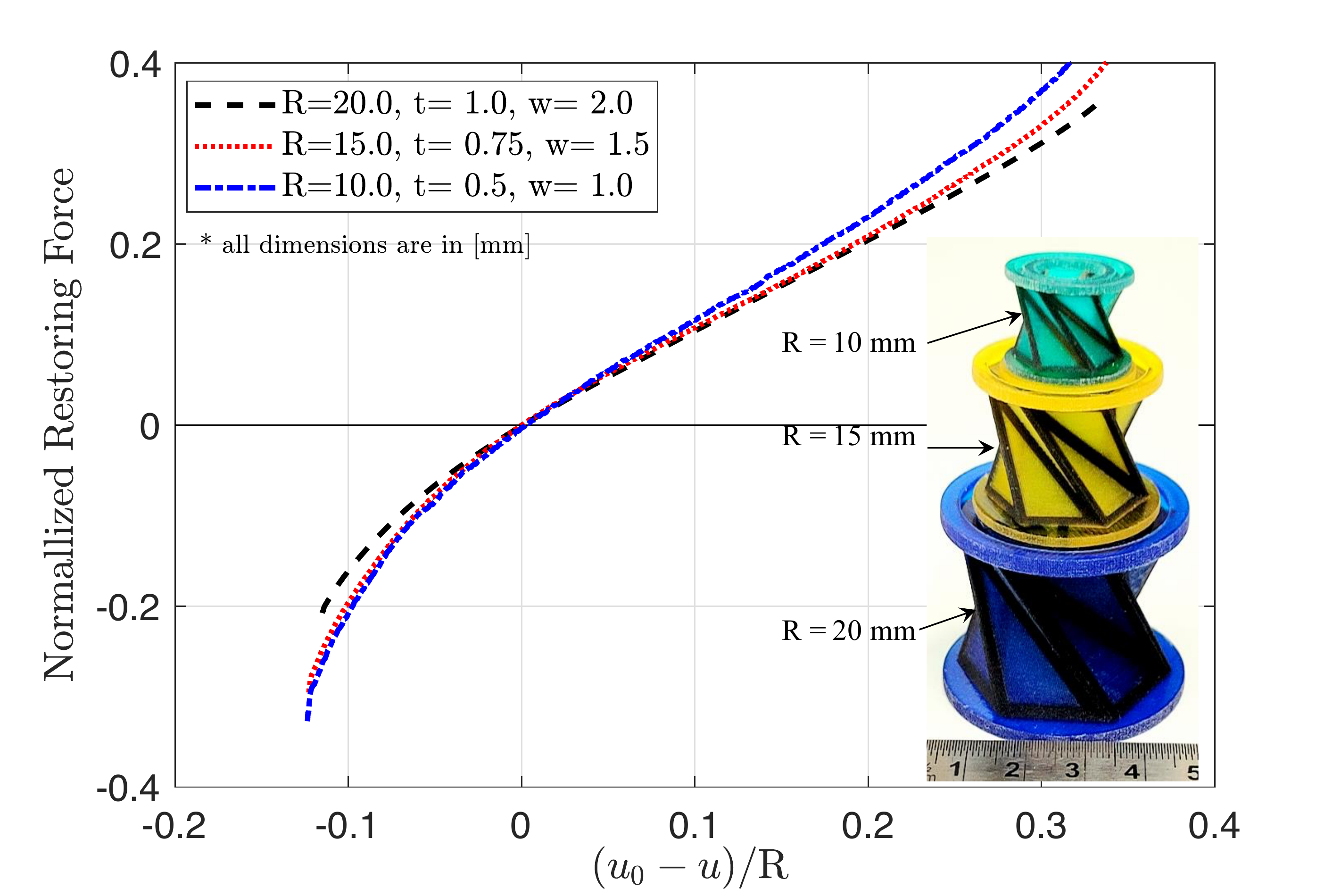}
		\caption{Normalized restoring force curves of a mono-stable KOS designed using the parameters, $u_0/R=1$, $\phi_0=90^o$ and $n=6$. Insets show the 3-D printed KOSs.} 
		\label{M1-scaled}
	\end{figure*}

\begin{figure*}[htb!]
		\centering 
		\includegraphics[width=12.0cm]{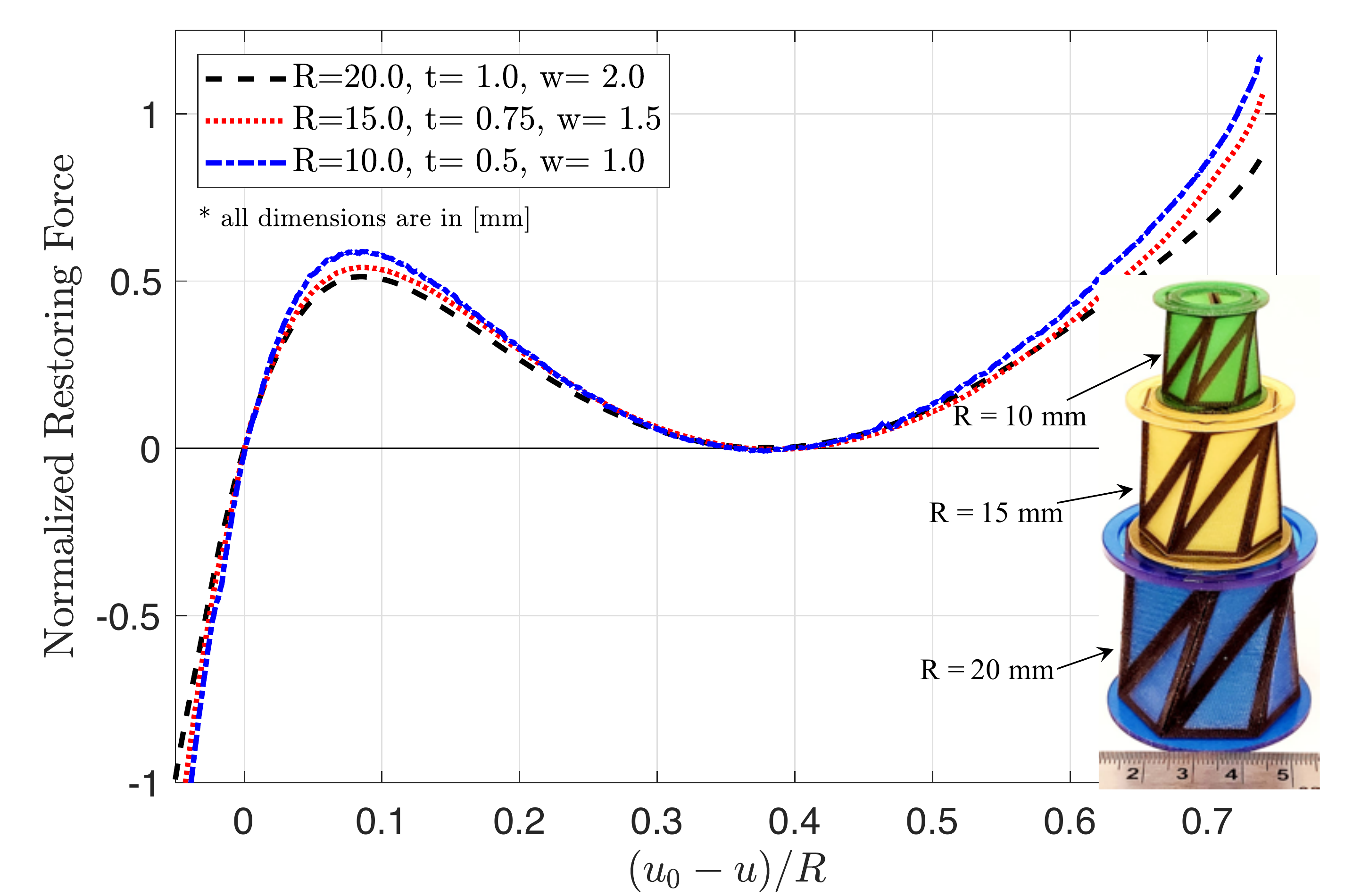}
		\caption{Normalized restoring force curves of bi-stable KOS designed using the parameters, $u_0/R=1.875$, $\phi_0=60^o$ and  $n=6$. Insets show the 3-D printed KOSs.} 
		\label{B1-scaled}
	\end{figure*}

It is already demonstrated that, by careful selection of design parameters, stiffer and softer springs can be designed. Yet, it is also important that a mechanical element like the spring be easily scaled depending on the stiffness requirements and physical space constraints. To assess the viability of scaling the KOS, different scaled versions of the KOS are fabricated and tested. To this end, all dimensions were scaled by the same amount such that the ratio between any dimension and $R$ remains constant.  Figures~\ref{M1-scaled} and \ref{B1-scaled}, respectively, show the restoring force curves for a mono- and bi-stable springs for three values of $R$; namely $R=20$ mm, $R=15$ mm, and $R=10$  mm. Here, the displacement was normalized by $R$, while the force was normalized by $kR$, where `$k$' is the empirical stiffness found in Equation~\ref{k-empirical}. It can be seen that restoring force curves of the different scaled KOS of a same design are qualitatively similar. Limitations on scaling down the KOS depend mainly on the capabilities of the 3-D printer, while limitations on scaling up  depend mainly on fabrication time and cost which scales with $R^3$.

\section{Stacking}
\label{stacking}
Another interesting feature of the KOSs is that they can be combined in parallel and series to either increase or decrease their effective stiffness. This feature was used in Ref. \cite{YasudaKresling2017} to demonstrate that KOSs can be used for mechanical memory bit operations. To explore such combinations, we fabricated two sets of KOSs, each consisting of two identical springs stacked in series. The first set  was fabricated using two identical mono-stable springs using the design parameters  $u_0/R=1.0$, $\phi_0=90^o$,  and $n=6$. The springs were stacked such that they have the same chirality in one test, and opposite chirality in the other. 

\begin{figure*}[htb!]
		\centering 
		\includegraphics[width=12.0cm]{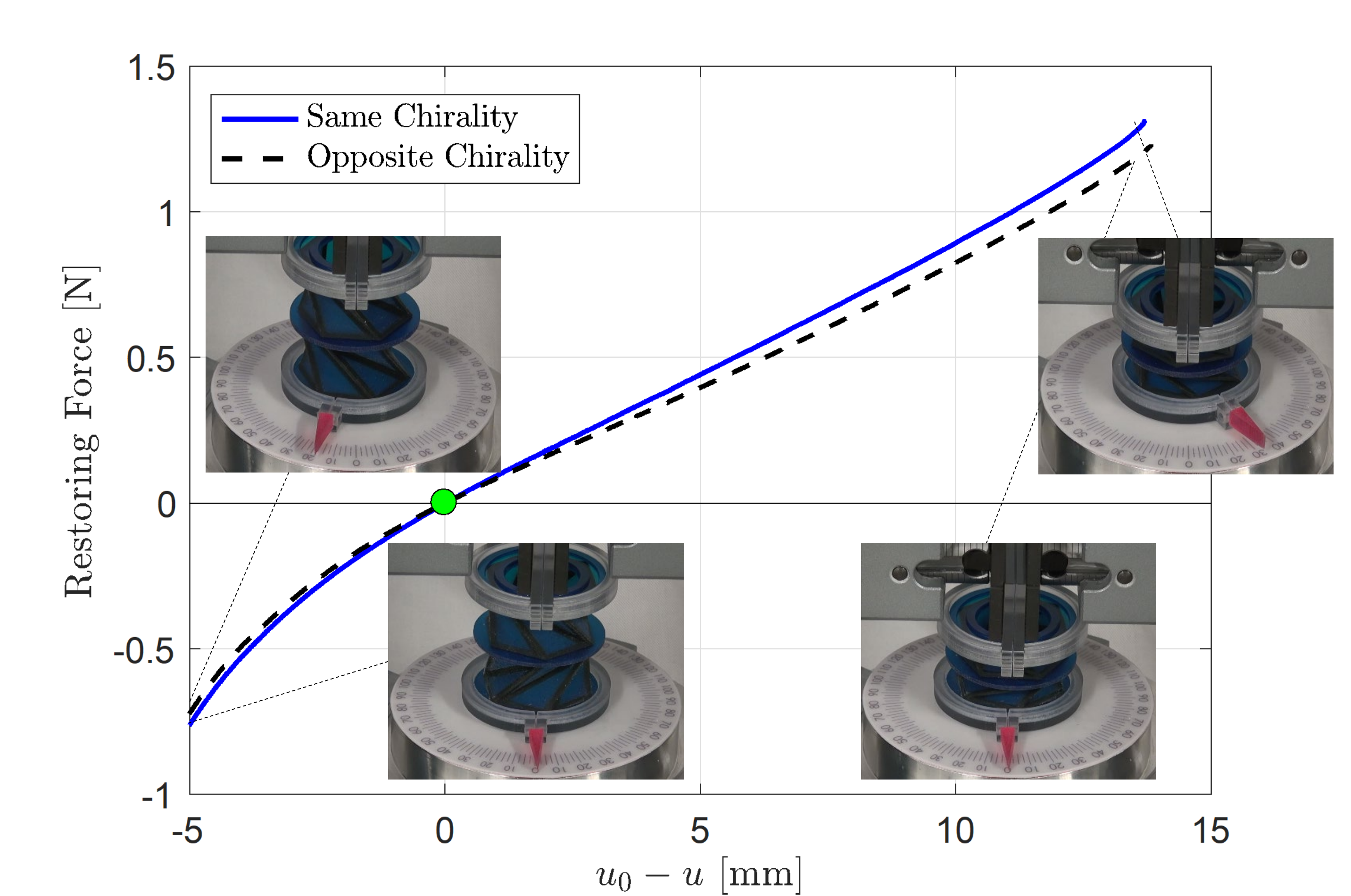}
		\caption{Restoring force curve for a stack of two identical mono-stable springs connected in series. Results are obtained with KOS designed using the parameters $u_0/R=1.0$, $\phi_0=90^o$, $n=6$, $R=20$ mm, $t=1$ mm and $w=2$ mm. Insets show 3-D printed KOS at different deployment heights.} 
		\label{M1_stack}
	\end{figure*}

As shown in Fig.~\ref{M1_stack}, the restoring force curves remain the same regardless of chirality and the linear stiffness of the two-spring stack is half that obtained using one spring as shown in Fig.~\ref{M1}. However, one key advantage of stacking the two springs such that they have opposite chirality is that the rotation of the two far ends relative to each other is totally removed. This is important when designing spring so as to eliminate torques at the support. Watch the video SI-2 in the supplementary files for compression tests in both scenarios.

\begin{figure*}[htb!]
		\centering 
		\includegraphics[width=12.0cm]{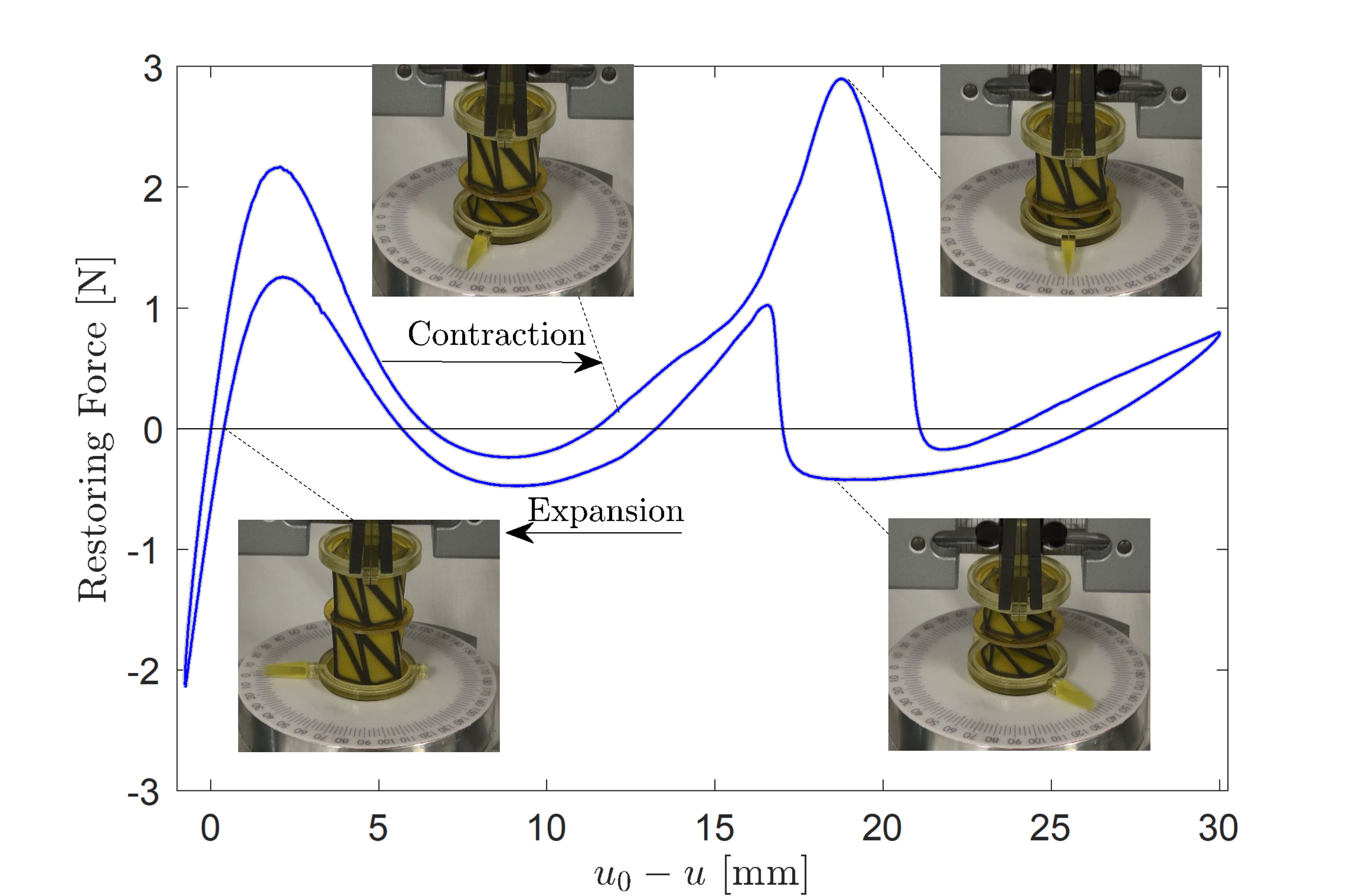}
		\caption{Restoring force curve for a stack of two identical bi-stable springs connected in series. Results are obtained with KOS designed using the parameters $u_0/R=1.65$, $\phi_0=45^o$, $n=6$,  $R=15$ mm, $t=0.75$ mm and $w=1.5$ mm. Insets show 3-D printed KOS at different deployment heights.} 
		\label{B2_stack}
	\end{figure*}

The second stack was fabricated using two identical bi-stable springs with each constructed using the design parameters:  $u_0/R=1.65$, $\phi_0=45^o$,  and $n=6$. Figure \ref{B2_stack} depicts the restoring force response for the stack, which illustrates that, although both springs forming the stack are identical, they do not compress simultaneously. As can be seen in the video SI-3 in the supplementary files, the lower KOS in the stack undergoes larger deformation initially and buckles first under the load, followed by the upper spring.

\section{Durability Analysis}
\label{Durability}

For any practical application, springs need to be durable and their responses repeatable even after many loading cycles. The materials we used for 3D printing the KOS are highly durable with a fatigue life of over $10^5$ cycles at a tensile strain of $20\%$ for \textit{Tango Black},  and over $10^4$ cycles for \textit{Vero} materials \cite{MooreFatigue2015}. The KOS design has a combination of stiffer  \textit{Vero}  and flexible \textit{Tango Black} that are not only subjected to tensile strains, but also to a complex combination of normal strain, shear strain, stretching and bending at the material interfaces. Such combinations  of loading are known to cause 3D printed materials to fail at the intrefaces \cite{VU2018447,LUMPE20191}.

\begin{figure*}[htb!]
		\centering 
		\includegraphics[width=12.0cm]{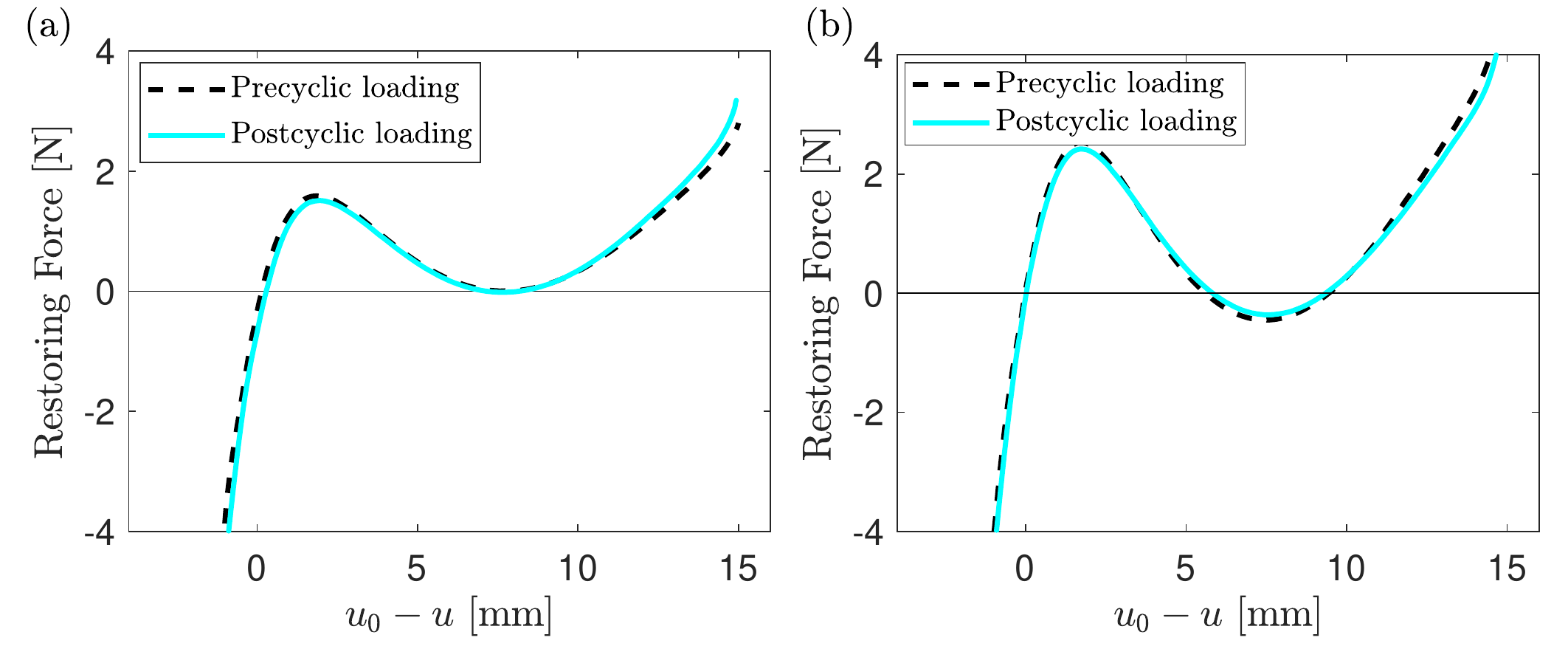}
		\caption{Restoring force curves before and after the fatigue tests of a bi-stable KOS designed using the parameters, $u_0/R=1.875$, $\phi_0=60^o$,  $n=6$, $R=20$ mm, $t=1$ mm and (a) $w=2$ mm, (b)  $w=1.5$ mm.} 
		\label{Fatigue}
	\end{figure*}

Towards assessing the durability of the KOSs, the KOS sample is first tested under quasi-static loading with a constant displacement rate of $0.2$ mm/s while the force-displacement response is recorded. Subsequently, the displacement rate is increased to $5$ mm/s and each speciment is subjected to a total of 5000  loading cycles over its whole deployment range. The quasi-static behavior is then recorded one more time and compared to the one obtained precyclic loading. Figure~\ref{Fatigue} depicts the quasi-static behavior of the KOS pre and post the 5000 loading cycles. It is evident that for the two designs considered, the cyclic loading did not produce any signifcant changes in the quasi-static behavior of the KOS which is indicative of its durability up to 5000 loading cycles.

\section{Conclusion}
\label{conclusion}
In this manuscript, we reported on the design, 3-D printing, and quasi-static behavior of functional and durable KOSs that mimic the qualitative behavior of paper-based KOSs. In the reported design, each of the fundamental triangles in the paper based-KOS is redesigned to accommodate for easy folding and stretching at the panel interfaces while providing sufficient rigidity such that it follows the Kresling origami functionality without collapsing under loading. Specifically, each triangular panel was redesigned using an inner central rigid core  and an outer flexible rubber-like frame. The proposed design was fabricated using Stratasys J750 3D printer implementing the polyjet technique, which involves depositing extremely thin layers of photopolymer and exposing each layer to UV light for immediate curing and hardening.

The general qualitative behavior of the fabricated KOSs was first analyzed using design maps generated by adopting a simplified axial truss model. It was shown that the adopted model can, for the most part, predict whether the resulting KOS will be mono- or bi-stable based on the chosen geometric design parameters. However, since the simplified model neglects the effect of panel self-avoidance, which occurs as the two parallel end planes are pushed closer to each other, the model can sometimes lead to fictitious equilibrium points that cannot be realized experimentally.

Quasi-static testing of a large set of fabricated springs demonstrated that the variance of the results among different samples is very small further pointing to the repeatability of the fabrication process. It also showed that, different variations of the springs' qualitative and quantitative behavior can be realized by simple changes in the geometric design parameters. In general, it was shown that KOSs with linear, softening, hardening, mono- and bi-stable behavior can be fabricated further pointing to the vast range of potential  applications of such springs in science and engineering.

\textbf{Acknowledgement}. This research was partially carried out using the Core Technology Platforms resources at New York University Abu Dhabi.

\textbf{Author Contributions}.
S.K. designed and conducted the experiments; R.M. analyzed the results and drafted the paper; M.D. conceptualized and edited the draft.


	\appendix

	
	\section{Fabrication of Paper KOS} \label{Fab_Con}	
		\setcounter{equation}{0}
	\setcounter{figure}{0}
	\renewcommand{\thefigure}{\thesection.\arabic{figure}}
	The paper based Kresling Origami Springs (KOS) were fabricated using 180 grams per square metre (GSM) paper. Depending on the desired potential charactersitics the design parameteres are selected from the design map, Fig.~\ref{Designmap}. Using an Epilog Fusion M2-40 laser cutting machine, very fine perforations were made on the paper in order to form the triangular panels and hanging B-flaps with holes shown in Fig.~\ref{KIMS_Fab} (Stage I). Creases were then made along the perforations and the paper was folded along the creases such that the shorter edges are folded outwards (a mountain fold) and the longer edges are folded inwards (a valley fold). The far end edges with A-flaps were then attached to the opposite end panels using glue to form a closed lamina (Stage II). The top and bottom surfaces of the KOS were subsequently reinforced using laser-cut acrylic polygon faces, which were attached to the flaps along the edges $a_0$ as shown (Stage III and IV). These plates, each weighing 22 grams, enable easy handling, and clamping of the KOS during quasi-static testing.
	
\begin{figure}[htb!]
	\centering 
	\includegraphics[width=12.0cm]{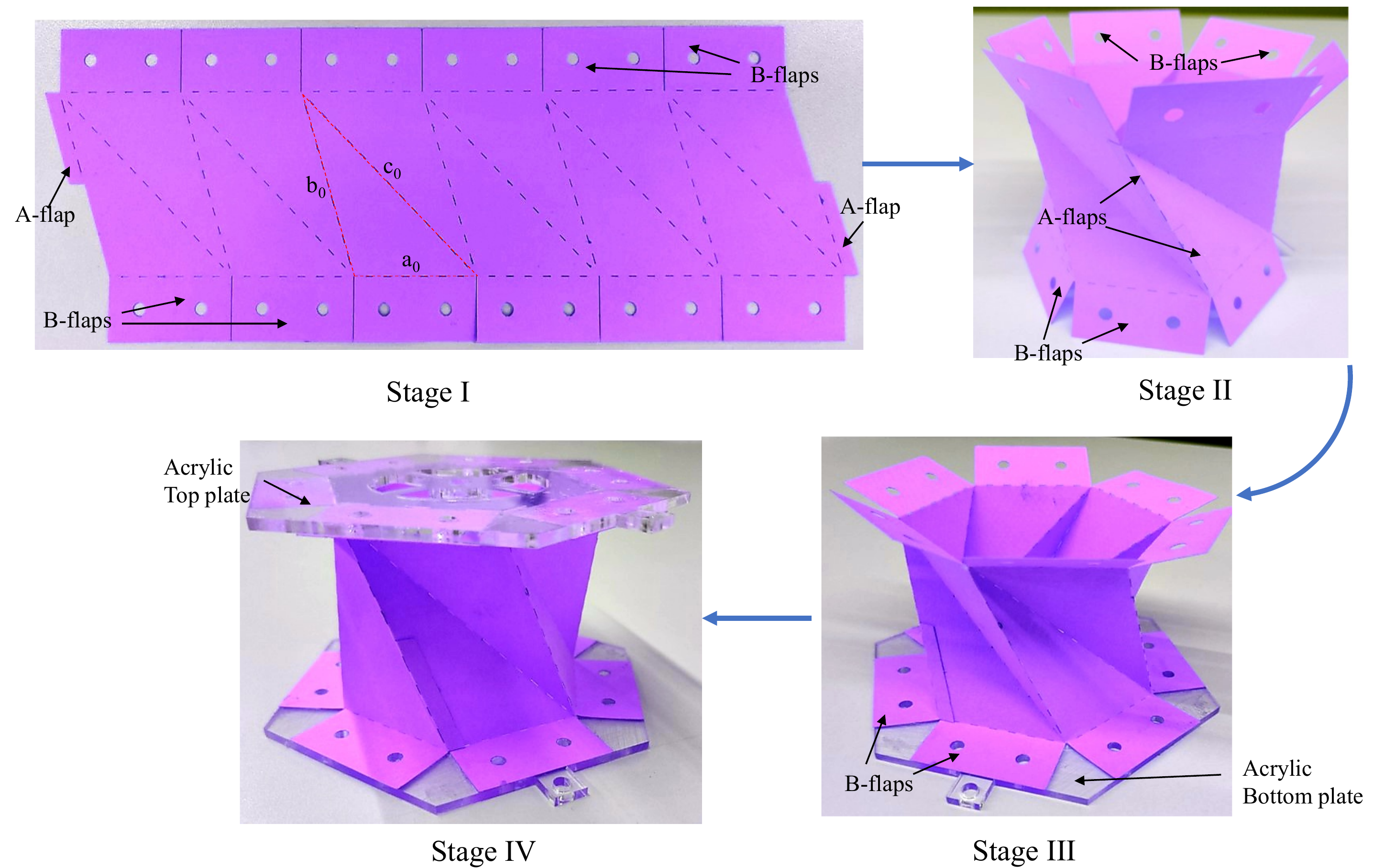}
	\caption{\protect\footnotesize Different stages of paper based KOS fabrication.  }
	\label{KIMS_Fab}
\end{figure}

\bibliography{KOS}

\end{document}